\documentclass[journal]{IEEEtran}

\usepackage{enumerate}
\usepackage[normalem]{ulem}
\usepackage{graphicx}
\usepackage{url}
\usepackage{amsthm}
\usepackage{comment}
\usepackage{balance}
\graphicspath{ {./images/} }
\theoremstyle{definition}
\newtheorem{definition}{Definition}

\ifCLASSINFOpdf
\else
\fi

\begin{document}
\title{An Introduction to Cyber Peacekeeping}

\author{Michael~Robinson, Kevin~Jones, Helge~Janicke and Leandros~Maglaras, Senior Member, IEEE
\thanks{M. Robinson is with Airbus, Newport, United Kingdom (michael.mi.robinson@airbus.com)}% <-this % stops a space
\thanks{K. Jones is with Airbus, Newport, United Kingdom (kevin.jones@airbus.com)}%
\thanks{H. Janicke is with the Software Technology Research Laboratory, De Montfort University, Leicester, United Kingdom (heljanic@dmu.ac.uk)}%
\thanks{L. Maglaras is with the Software Technology Research Laboratory, De Montfort University, Leicester, United Kingdom (leandrosmag@gmail.com)}}%

\markboth{April~2018}%
{Shell \MakeLowercase{\textit{et al.}}: An Introduction to Cyber Peacekeeping}

\maketitle

\begin{abstract}
Cyber is the newest domain of war, and the topic of cyber warfare is one that is receiving increasing attention.  Research efforts into cyber warfare are extensive, covering a range of issues such as legality, cyber weapons and deterrence.  Despite all of the research activity around cyber warfare, one aspect has been largely overlooked: the restoration of peace and security in its aftermath.   In this article, we present the argument that cyber warfare will threaten civilian peace and security long after a conflict has ended, and that existing peace operations will be required to evolve in order to address this threat.  We explore how existing UN peacekeeping operations could be adapted, in ways that would be both feasible and valuable towards maintaining and restoring peace in a region.  We conclude that the path to cyber peacekeeping will not be easy, but that it is an evolution that must begin today so that we can be prepared for the conflicts of the future.
\end{abstract}


% \fi
\IEEEpeerreviewmaketitle

\section{Introduction}
% The very first letter is a 2 line initial drop letter followed by the rest of the first word in caps.
\IEEEPARstart{T}{he} topic of cyber warfare continues to receive a great deal of coverage in the media~\cite{Weinberger2013,Goldman2013,Gelzis2017}, on the political stage~\cite{Dominiczak2014,Fisher2013} and in academia~\cite{Chen2010,baradaran2017}.  The research community has responded to the rise of cyber warfare by taking on some of the most pressing and immediate challenges that the topic presents: the legality, ethics and doctrine of cyber warfare have all been debated in great detail, along with other important topics~\cite{Robinson2014}.  While research into cyber warfare is clearly extensive, there remains an area where research is lacking: how do we restore peace in the aftermath of cyber warfare?  We address this question by considering the concept of cyber peacekeeping.

Section~\ref{backgroundsection} sets out the background to the topic.  We begin by discussing what is meant by the term peacekeeping and adopt the UN definition.  We ask the question of whether cyber peacekeeping is necessary, analysing arguments both for and against its existence.  Previous work on the topic is discussed and definitions of both cyber peacekeeping and cyber peacekeepers are proposed.  We then examine each existing UN peacekeeping activity for both value and feasibility in a post-cyber warfare context.  The article concludes with directions for future research and the conclusions of our research.

\section{Background}\label{backgroundsection}
In order to consider cyber peacekeeping, we must first be clear what is meant by both cyber warfare and peacekeeping.

\subsection{Cyber Warfare}
Cyber warfare is a vast area of research, with a number of fields of study.  Research topics can range from the ethical considerations of cyber warfare~\cite{Taddeo2012} to how cyber weapons could be developed and used~\cite{Tyugu2012}.  There is however little consensus regarding many aspects of cyber warfare, including how it should be defined~\cite{Robinson2014}.  Despite this lack of understanding, a number of national militaries now regard cyber space as a warfighting domain, and have developed doctrines for fighting within it~\cite{dod2013b,connell2017}.  This lack of understanding combined with increasing state level tensions over cyber conflict~\cite{Sengupta2018} are a concerning situation:  we are pioneers in a new and untested form of conflict where the boundaries are unknown, international law is yet to develop and attribution for actions is difficult.

% The definition below has now moved to here, from the definition section.  It makes sense to have it this section, along with the definition we adopt of peacekeeping.  This then leaves the definitions section to focus solely on cyber peacekeeping.
Although we have not reached international consensus on how cyber warfare should be defined, it is necessary to adopt a definition prior to considering a concept such as cyber peacekeeping.  Therefore, where the terms cyber attack, cyber war and cyber warfare are used in this article, the definitions proposed by Robinson et al.~\cite{Robinson2014} are adopted. These definitions are as follows:

\begin{itemize}
	\item Cyber Attack: An act in cyber space that could reasonably be expected to cause harm.
	\item Cyber Warfare: The use of cyber attacks with a warfare-like intent.
	\item Cyber War: Occurs when a nation state declares war, and where only cyber warfare is used to fight that war.
\end{itemize}

\subsection{Peacekeeping}
The term peacekeeping is defined by the Oxford English Dictionary as ``The active maintenance of a truce between nations or communities, especially by an international military force''~\cite{Oxford2014}, but many definitions exist with little international agreement on which is correct~\cite{Bellamy2010}.

We therefore adopt the definition used by one particular organisation:  The United Nations (UN).  Arguably the most high profile peacekeeping organisation of today, the UN receives extensive media coverage which exposes their work to the public~\cite{Sheridan1993, Blystone1995}.  The UN definition of peacekeeping is as follows: ``Action undertaken to preserve peace, however fragile, where fighting has been halted and to assist in implementing agreements achieved by the peacemakers''~\cite{UN2008}.  While we have picked UN peacekeeping as our focus, it must be noted that many other organisations perform peacekeeping.  The Multinational Force and Observers (MFO) is one such example of a non-UN organisation whose stated aim is to undertake peacekeeping responsibilities in the Sinai~\cite{MFO2014}.

%new subsection because we specifically discuss UN Peacekeeping.
\subsection{UN Peacekeeping}
To understand UN peacekeeping today, we must have an understanding of its origins and history.  UN peacekeeping was developed as a means for the UN to meet one of its core purposes as defined in the UN Charter: to maintain international peace and security.

The first United Nations peacekeeping operation was formed in 1948, in response to concern over hostilities in the Middle East.  The United Nations Truce Supervision Organization (UNTSO) was dispatched as an unarmed monitoring mission to monitor the Armistice Agreement between Israel, Egypt, Lebanon, Jordan and Syria~\cite{Macqueen1999}.  Throughout the 1960s and 70s, the UN continued to dispatch small scale peacekeeping operations, helping to monitor ceasefires and uphold peace in post conflict environments.

Into the 1980s, the ambition of UN peacekeeping continued to grow both in scope and numbers.  New operations went beyond the traditional peacekeeping tasks seen previously, expanding in scope to include more complex goals such as the supervision of elections, providing humanitarian support and building democratic institutions.

While earlier operations were generally regarded as successful, it was during this period that the UN started to encounter high profile failures.  The Bosnian War of 1992 and the UN Operation in Somalia II (UNOSOM II) in 1993 are both regarded as failings in UN peacekeeping~\cite{roberts1994}.  In Somalia, UN peacekeeping troops effectively became a participant in the conflict rather than acting as peacekeepers.  In Bosnia, UN peacekeepers failed to prevent ethnic cleansing and ensure the protection of civilians.  The Dutch state was later found liable for 300 deaths, with The Hague district court ruling that Dutch peacekeepers did not do enough to ensure their protection~\cite{spijkers2016}.  Similarly in the high profile case of Rwanda (1994), UN peacekeepers failed to protect civilians from harm~\cite{barnett1999}.

These failures led the UN to re-evaluate its approach to peacekeeping~\cite{UN201450years}, and a number of reforms were recommended~\cite{Brahimi2003}.  This resulted in the realisation that peacekeeping was not a magic wand.  It had limitations and was only appropriate when specific criteria were met.

Today, UN peacekeeping still faces multiple challenges.  As an example, MONUSCO is the most expensive ongoing peacekeeping operation in history.  Aimed at restoring peace in the DR Congo, it was established in 1999 and is still ongoing today (October 2017) with 22,000 uniformed personnel deployed in the region and a budget of just over \$1b per year.  Despite the long involvement of peacekeeping forces and some important successes such as the holding of free and fair elections in 2006, fighting between combatants and attacks on civilians continues and there is little sign of long term peace.  Other problems facing UN peacekeeping include a constant struggle to secure troop and police contributions, weak support from the international community and the internal challenge of running such a huge organisation effectively.

\subsection{UN Peacekeeping Goals}\label{goals}
The UN defines two type of peacekeeping operation that it can undertake:  Traditional and Multi-Dimensional.

\begin{itemize}
  \item \textbf{Traditional}: Operations which adhere strictly to the traditional goals of observation, monitoring and reporting.
  \item \textbf{Multi-Dimensional}: More complex operations which include peacekeeping but also extend into peace building e.g. reforming a state's security sector and clearing mines.
\end{itemize}

\subsection{Peacekeeping amongst other activities}
Whilst this article is focused upon cyber peacekeeping, it is important to note that peacekeeping is an activity which overlaps with a wider set of peace operations.  According to the UN, these other activities are:

\begin{itemize}
  \item \textbf{Conflict Prevention}: Early intervention to prevent a dispute escalating.
	\item \textbf{Peacemaking}: Diplomatic measures aimed at bringing about a ceasefire.
	\item \textbf{Peace Enforcement}: Restoring peace without consent of the parties.
	\item \textbf{Peace Building}: Laying the foundation for long term peace and preventing relapse into conflict.
\end{itemize}

\begin{figure}[ht]
\centering
\includegraphics[width=\columnwidth]{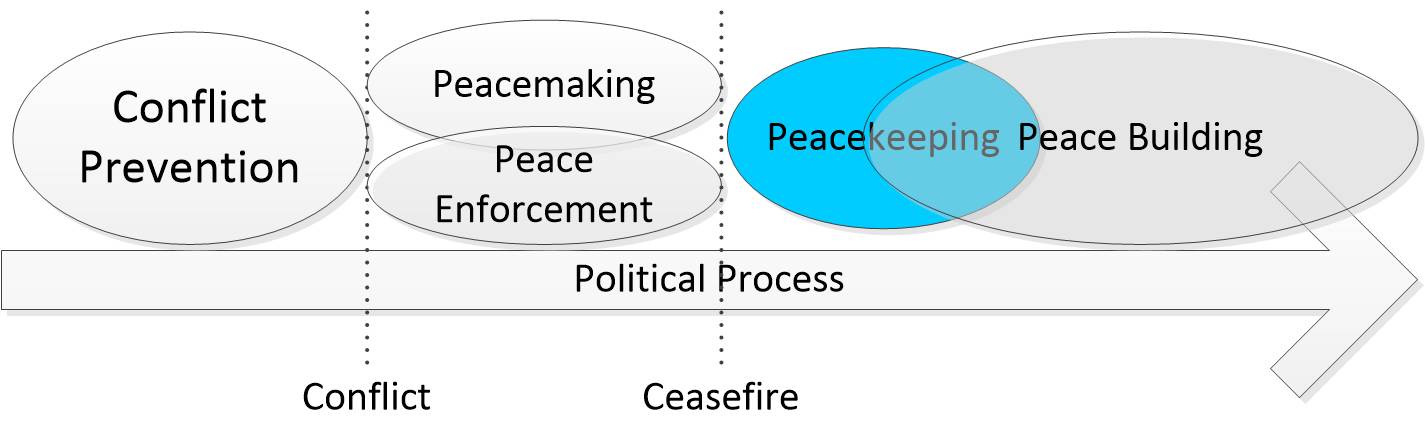}
\caption{UN Peace Activities~\cite{UN2008}}
\label{fig:peaceactivities}
\end{figure}

These activities and their relationship to peacekeeping are shown in figure \ref{fig:peaceactivities}.

The overlap between each of the activities is clear; Peacekeeping operations cannot be viewed in isolation, and peacekeepers are often called upon to assist in peacemaking and peace building where necessary.

\subsection{UN Peacekeeping Principles}
UN Peacekeeping has always been guided by a number of core principles, further refined following the Brahimi report~\cite{Brahimi2003}.  Today, these core principles are as follows:

\begin{itemize}
  \item \textbf{Consent of the parties}: Peacekeeping operations are only deployed with the consent of the conflicting parties.  This gives the operation the legitimacy to act both physically and politically in the area.  Without consent of all parties, the operation risks becoming involved in the conflict.
  \item \textbf{Impartiality}: Operations maintain peace without favouring any of the involved parties.  An operation must be seen as impartial to remain legitimate.  Following the Brahimi report, it was clarified that impartiality does not mean inaction in the face of clear threats to peace.
  \item \textbf{Non use of force, except in self defence and defence of the mandate}: Use of force should be a measure of last resort.  Following the Brahimi report, it was made clear that force could be used in defence of the mandate.  This highlighted that force could be used against those who were determined to undermine the peace process.
\end{itemize}

With this understanding of UN peacekeeping's history, purpose and principles, we are better equipped to begin exploring the concept of cyber peacekeeping.  The following section begins this process, by surveying existing works on the topic.

\section{Existing Research}\label{previouswork}
One of the earliest explorations into the concept of cyber peacekeeping came from Cahill et al.~\cite{Cahill2003}, who identified cyber peacekeeping as a future area of research. They examined existing UN peacekeeping principles and made proposals on how they could be applied to the cyber domain.  The work by Cahill et al. is valuable in that it made the first steps towards identifying cyber peacekeeping as a novel area of research, and followed a logical approach towards proposing the six principles. While their work lays good foundations for thinking about cyber peacekeeping, the authors did not publish further work to expand upon their proposals.  This is a problem, since there are significant questions regarding how each principle could be implemented at a practical level.

Research by Kleffner and Dinniss~\cite{Kleffner2013} concurs that future conflict will contain cyber elements, and that “peacekeepers will increasingly find themselves on missions in which cyber incidents will occur during, following or even in the absence of, conventional hostilities.”  Their focus was to examine cyber peacekeeping from a legality perspective and they state that the UN Security Council would be legally entitled to determine that cyber warfare could amount to a threat to international peace and security under Article 39 of the UN Charter. They conclude that cyber peacekeeping will largely be legally permissible, with international human rights law and the specifics of the mandate for a particular operation being the primary legal framework to guide it.

The most recent work regarding cyber peacekeeping was contributed by Akatyev and James~\cite{Akatyev2015}  . They propose that cyber peacekeeping has a role during all three stages of a conflict: pre-conflict, conflict and post-conflict.  In the pre-conflict stage, cyber peacekeepers will conduct activities to maintain and enhance international peace. For example, working to develop international norms and contributing to research and development. During the conflict stage, the tasks change. Cyber peacekeepers now work to detect attacks and minimise their impact on civilians while assisting nations in keeping critical infrastructure operational. In the post conflict stage, they work to help states develop countermeasures to the cyber weapons used in the conflict, while also assisting nations in rebuilding their defences. They describe two practical implementations of cyber peacekeeping - a Rapid Response Division (RRD) and a Long Term Stability and Relief Division (LSRD). The focus of the RRD is primarily to protect what they term the “cyberspace safe layer”. This is defined as being a nation’s “pre-identified, minimally-required critical infrastructure necessary for civilian safety”. In the case of cyber conflict, the RRD will therefore take immediate measures to protect and ensure the availability of this safe layer. The LSRD takes a longer term approach, working to build capacity and defences to ensure a longer lasting peace.  Akatyev and James followed up their research in 2017\cite{Akatyev2017}, by proposing how their system could fit into new schemes such as the UN’s “digital blue helmet” programme.

\section{The Need for Cyber Peacekeeping}\label{need}
The use of the cyber domain is a relatively new aspect of war~\cite{Robinson2014}, and the world has yet to see a war in which cyber warfare has played a significant part.  We must therefore ask the question: is cyber peacekeeping necessary? 

As a starting point, let us consider why any form of UN peacekeeping is needed.  Chapter I of the UN Charter~\cite{UNCharter} defines the core objectives of the UN, one of which is to maintain international peace and security.  Peacekeeping operations are the practical measure through which the UN fulfils this mandate, and is the most common justification for forming a peacekeeping operation~\cite{Bellamy2010}.  If cyber warfare could threaten international peace and security, it follows that the UN would be required to conduct peacekeeping in response to it.  We must therefore consider if cyber warfare could realistically threaten international peace and security.  The answer to this depends upon how the term is defined.  When the term was originally used in the UN Charter, it was envisioned that most cases would arise out of inter-state conflict~\cite{Bellamy2010}.  Over the years however, the UN has widened the scope to include events such as state collapse, HIV/AIDS, nuclear proliferation, humanitarian suffering and massive human rights abuses.  In effect, the UN Security Council has a large degree of discretion on what constitutes a threat to international peace and security.

If events such as state collapse, humanitarian suffering and human rights abuses are threats to international peace and security, it is an indicator that cyber peacekeeping will be necessary.  Cyber attacks can initiate, compound or prolong such events.  For example, cyber attacks could at least contribute towards the collapse of a state if they initiate or prolong the failure of critical national infrastructure.  Nations are becoming reliant on the cyber domain to provide services that keep a nation running: power grids, water supplies, communications, transportation and finance are all increasingly becoming cyber dependant~\cite{cornish2010}.  Cyber warfare which causes blackouts, cuts off supplies to safe drinking water, makes travelling dangerous or destabilises a national economy is clearly a threat to the stability of a nation and is therefore a threat to international peace and security, providing justification for the establishment of a peacekeeping operation.

A similar argument can be made for human rights violations.  At the most critical end of the spectrum, a national blackout or toxic water supply has the potential to threaten the right to life.  Less grave but still important is the right of every person to seek, receive and impart information and ideas through any media and regardless of frontiers~\cite{UDHR}.  As an example, when the Cameroon government cut off internet access to predominantly English speaking parts of the country, the UN stated that it was ``an appalling violation of their right to freedom of expression''~\cite{UNOG2017}.  This evidences the fact that the UN believes human rights can be threatened in the cyber domain and therefore adds weight to the argument that cyber could amount to a threat to international peace and security.

We must also consider softer forms of cyber warfare such as election hacking as a justification for cyber peacekeeping.  Free and fair elections are an essential ingredient towards peace in a nation~\cite{UNElections2007}: any doubt over the legitimacy of an election result can damage a peace process and precipitate to a relapse into conflict.  Alleged targeting of elections and political systems by cyber means is becoming more common, with extensive media coverage over possible Russian influence on the 2016 US presidential election~\cite{Shane2017}.  Similarly, a 2017 cyber attack on Emmanuel Macron's party in France was described as a clear attempt to destabilise the election and democratic process~\cite{Willsher2017}.  These attacks were upon nations which are stable enough to withstand them.  However, in smaller nations recovering from conflict where peace is fragile, similar doubt upon election legitimacy could be enough to derail the peace process.

Looking to existing literature, opinion is divided.  Kleffner and Dinniss~\cite{Kleffner2013} have suggested that cyber peacekeeping will become a necessity, predicting that peacekeepers will find themselves having to operate inside of the cyber domain in order to maintain peace in future conflicts.  Akatyev and James~\cite{Akatyev2015} agree, stating that cyber peacekeeping is needed to protect an increasingly-connected number of people, to help prevent escalation of cyber conflicts, to provide arbitration among states, and to help build and maintain trust in cyberspace.  John Bumgarner, Chief Technology Officer at the U.S. Cyber Consequences adds weight to the argument for cyber peacekeeping by stating that ``the UN needs to figure out how they can deploy peace keepers in the digital borders of a nation, virtual peacekeepers that would protect the peace''~\cite{Watts2012}.

On the opposing side of the debate, others have argued that cyber peacekeeping is not necessary since it is both premature and redundant~\cite{Phneah2012}.  The premature argument proposes that until cyber warfare is better understood and more of a direct threat to society, attempts to design cyber peacekeeping are premature:  we simply do not understand the threats and how to counter them effectively.  This view can be challenged however, with a counter argument that cyber peacekeeping is necessary because cyber warfare is not fully understood~\cite{Robinson2014}.  As nations begin to use the cyber domain during warfare, unregulated, untested and experimental forms of cyber attack have the potential to unintentionally inflict indiscriminate, disproportionate and prolonged suffering to civilians.  This clearly presents a threat to international peace and security through human rights violations and state collapse.

A further lens through which a need for cyber peacekeeping can be considered is the concept of “responsibility to protect” (R2P).  R2P is an international agreement that all states have a responsibility to protect civilians from genocide, war crimes, ethnic cleansing and crimes against humanity, and was agreed upon unanimously by world leaders in 2005~\cite{Cooper2009}.  Although R2P can be used as a justification for peacekeeping, its justification for cyber peacekeeping is questionable.  While cyber warfare can target critical infrastructure and result in harm to civilians, it is questionable whether this activity would rise to the level of a war crime or genocide.  However, in the case of attacks on civilian air traffic control, water supplies, or dams the potential for indiscriminate and mass civilian casualties is high.  Therefore the potential for war crimes via cyber warfare and the relevance of R2P cannot be entirely dismissed.

When considering a political construct such as peacekeeping, it is also necessary to consider political views in relation to whether cyber peacekeeping is needed.  For example, the Westphalian school of thought believes that the world consists of sovereign states who recognise no higher authority.  Other states and organisations should not interfere with issues inside of that state unless invited~\cite{Bellamy2010}.  If approached from this perspective, cyber peacekeeping is harder to justify, but still possible when full consent from both parties is given - a core principle of UN peacekeeping.

In summary, while there are arguments both for and against the need for cyber peacekeeping, the balance of arguments favour cyber peacekeeping becoming necessary in the future.  These are summarised into three points:

1.	Cyber warfare is new and untested, raising the likelihood that it could unintentionally lead to indiscriminate, disproportionate and prolonged suffering to civilians, even after a conflict has ended.
2.	Cyber warfare could threaten international peace and security in a variety of ways, satisfying the UN Charter’s own requirements for the establishment of a peacekeeping operation.
3.	Some cyber warfare might rise to a level which exceeds the thresholds of responsibility to protect (R2P).

\section{Defining Cyber Peacekeeping}\label{CPKingdefinition}
We have put forward the argument that cyber peacekeeping will become necessary in the future, and it is now important to define what is meant by “cyber peacekeeping”.  This is not a simple task, since the question of how to define “peacekeeping” even in its current form is a source of continual debate~\cite{Bellamy2010}.  To simplify the task, we adopt the UN definition:

\begin{quote}
``Action undertaken to preserve peace, however fragile, where fighting has been halted and to assist in implementing agreements achieved by the peacemakers.''~\cite{Mays2010}
\end{quote}

As a first step towards finding a definition, the original can be modified to focus it around cyber space:

\begin{quote}
``Action undertaken \textit{in cyberspace} to preserve peace, however fragile, where fighting has been halted and to assist in implementing agreements achieved by the peacemakers.''
\end{quote}

The primary strength of this definition is that it is built upon wording that is already established in the international community.  A potential weakness is that requiring an action be performed “in cyberspace” to count as cyber peacekeeping means that activities which require cyber knowledge but are kinetic in nature (such as training or assisting with policy reforms) cannot be regarded as cyber peacekeeping. To address this weakness, an alternative definition can be considered:

\begin{quote}
\textit{Cyber related} action undertaken to preserve peace, however fragile, where fighting has been halted and to assist in implementing agreements achieved by the peacemakers.
\end{quote}

This definition removes the requirement that the action must be in cyber space, but to determine which definition is correct an important question must be asked: should the test for whether an action is cyber peacekeeping be the type of knowledge being used or the actual domain it is being performed in? For assistance here, we can study a similar case where peacekeeping has been applied to a particular domain of war: peacekeeping in the maritime domain. The UN Peacekeeping Missions Military Maritime Task Force Manual~\cite{UNMaritimeManual2015} states the following:

\begin{quote}
``The Maritime Task Force is not limited to maritime effects at sea.  Its ships provide advanced platforms for military aviation, communications and medical support to the ground force.  These maritime capabilities strengthen the Force Commander's capability with enhanced deterrence, situational awareness, medical facilities and military transportation sustaining operations and the execution of mandated tasks on shore and afloat.''~\cite{UNMaritimeManual2015}
\end{quote}

This suggests that a naval task force dispatched to conduct ``maritime peacekeeping'' is not confined to conducting tasks at sea.  The reference to maritime capabilities being used to strengthen the operation suggests that naval peacekeeping is defined by the application of a \textit{maritime capability}, rather than the action being confined to any specific domain.  This proposal is supported further by the manual, which goes on to describe training in areas such as maritime law and policy as being a task of the maritime task force.  It is therefore argued that if maritime peacekeeping is the application of maritime capability in any domain, it follows that cyber peacekeeping should be the application of cyber capability in any domain.  The definition can therefore be refined as follows:

\begin{definition}{UN Cyber Peacekeeping.}
The application of cyber capability to preserve peace, however fragile, where fighting has been halted and to assist in implementing agreements achieved by the peacemakers.
\end{definition}

For completeness, it is also necessary to define the term cyber peacekeeper.  The term `peacekeepers' has been defined as ``individuals and groups who perform peacekeeping activities''~\cite{Bellamy2010}.  This is a broad definition, and therefore includes civilians, armed forces, charities, governmental and non-governmental organisations. The UN does not provide a definition of a peacekeeper.  Whether the term refers to uniformed blue helmets, or to all people involved in a UN peacekeeping operation such as civilian staff, volunteers and partner organisations remains unclear.  We adapt the definition offered by Bellamy and Williams~\cite{Bellamy2010} as follows:

\begin{definition}{Cyber Peacekeeper.}
An individual performing cyber peacekeeping activities.
\end{definition}

\section{Methodology}\label{activities}
We identified two potential approaches towards developing the activity of cyber peacekeeping.  Firstly, it could be designed as a unique concept with a set of activities drawn up from scratch.  Secondly, we could examine existing UN peacekeeping activities and explore how they could be adapted to bring value in a post-cyber warfare context.  We propose that the latter approach brings a number of benefits:

\begin{itemize}
  \item \textbf{Adoption}: When cyber peacekeeping can be shown to fit alongside established peacekeeping norms, peacekeeping organisations will be more likely to understand the value and carry it forward into the field.
  \item \textbf{Comprehensive}: Peacekeeping activities have been developed over decades, in response to practical experience in the field.  By adapting existing activities, it is more likely that cyber peacekeeping will address the issues that are significant to peacekeeping operations.
	\item \textbf{Integration}: By sharing a common approach, cyber peacekeeping has the flexibility to either operate as a standalone event, or alongside kinetic peacekeeping as part of a wider operation.  Cyber peacekeepers can be smoothly integrated into existing processes, where overlap with the other domains will be common.
\end{itemize}

Taking this approach, we adopt the following methodology in this article:

\begin{enumerate}
	\item Each existing UN peacekeeping activity is presented, and the value that it brings towards maintaining peace identified.
	\item We consider how this value could be maintained if the activity was performed in relation to cyber warfare.   If no value can be identified, even by adapting the activity, it should not be included as part of cyber peacekeeping.
	\item If some value can be maintained, we will explore how feasible the activity would be to perform in a cyber warfare context.  Where obstacles are expected, these will be discussed.
	\item The expected result is the proposal of a collection of cyber peacekeeping activities which are familiar to peacekeeping organisations, feasible to perform and valuable towards maintaining international peace and security.
\end{enumerate}

\begin{figure*}[ht]
	\centering
	\includegraphics{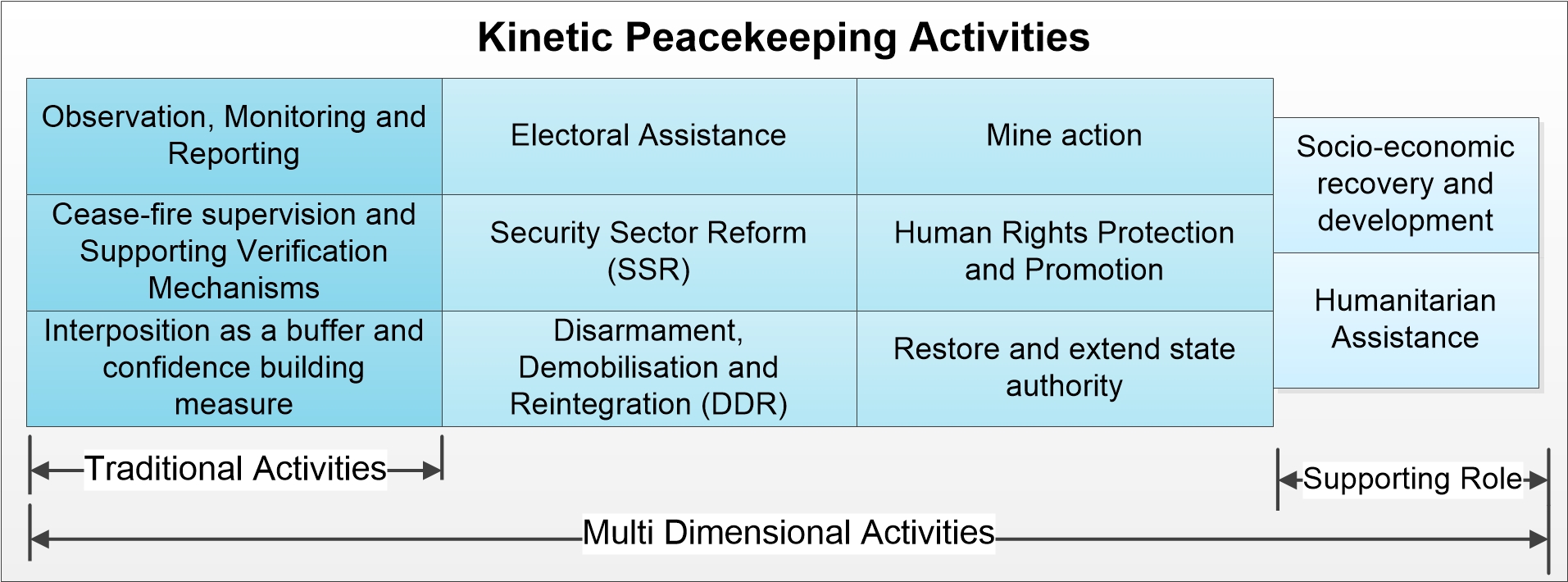}
	\caption{Activities undertaken by kinetic UN peacekeeping operations as defined by the UN Department of Peacekeeping Operations (DPKO)}
	\label{fig:kineticactivities}
\end{figure*}

UN peacekeeping literature~\cite{UN2008} indicates that there are broadly eleven distinct activities that are performed in a UN peacekeeping operation.  These are shown in figure~\ref{fig:kineticactivities}.

In line with our methodology, we now examine each activity and identify how each one brings value towards maintaining peace and security.  We consider ways in which that value could be transferred to a post-cyber warfare context.  If value can be found, we then consider how feasible it would be and discuss any challenges that might arise.

\section{Observation, Monitoring and Reporting}
Observation, Monitoring and Reporting (OMR) is one of the core activities performed during a UN peacekeeping operation.  The activity brings value by providing impartial reporting on adherence to ceasefire agreements, violations of human rights and other information where trust in its correctness is critical towards maintaining peace.  OMR can be a specific task, such as a foot patrol around a specific area or by the dispatch of unarmed UN Military Observers (UNMOs).  It is also an incidental task, performed by peacekeeping staff who observe and report issues during the course of their other peacekeeping duties.  UN Infantry Battalion Manuals Volumes One and Two~\cite{UNIBAM1,UNIBAM2} specify three observational goals of OMR.  UN personnel aim to observe, monitor and report upon:

\begin{enumerate}
	\item Actions which violate peace agreements.
	\item Human rights abuses.
	\item Changes in terrain, dispositions and civilian activity.
\end{enumerate}

When considering how OMR could apply to the cyber domain, an immediate point of note is that the concept of observation, monitoring and reporting in cyberspace is not new.  Cyber attacks against businesses and governments have encouraged research into how observation, monitoring and reporting in cyberspace can be improved~\cite{Chen2007,Ye2004}.  Books by authors such as Bejtlich~\cite{Bejtlich2004,Bejtlich2013} and Murdoch~\cite{Murdoch2014} provide advice on best practice regarding network monitoring and reporting, and certifications such as the Cisco cybersecurity specialist certification~\cite{Cisco2015} are designed to give cyber security professionals the skills needed to effectively observe and monitor in the cyber domain.  It is therefore true that many of the existing techniques of monitoring in cyberspace will likely be used by cyber peacekeepers.  The focus of this section is therefore not to discuss how cyberspace can be monitored and observed, since this is an already established field of study. The focus is to explore how existing cyber observation methods could be used to fulfil the three OMR observational goals.  We begin by examining the first goal.

\subsection{Actions which violate peace agreements}
The monitoring and reporting of actions which violate peace agreements brings value towards maintaining peace, since the reports will be coming from a trusted third party which is unbiased towards any particular side to the conflict.  This means that the information reported by UN peacekeepers can be regarded as true and fair, removing doubt and potential accusations of false reporting.  While clearly valuable in relation to kinetic warfare, can this value be maintained when applied to cyber warfare?

It has been stated that wars of the future will contain elements of cyber warfare~\cite{Robinson2014}.  If this is true, it follows that future peace agreements will likely contain terms relating to cyber warfare.  At the very least, it is reasonable to expect a peace agreement to state that cyber attacks should cease.  If cyber peacekeeping can monitor and report on violations of these terms in an unbiased and fair manner, then value can be found.  For example, if country A agrees not to attack country B's power grid, UN cyber peacekeeper monitoring would be valuable in acting as a trusted third party to monitor for adherence to that agreement.  It is therefore prudent to explore whether it would be feasible to perform.

\begin{figure*}[ht]
\centering
\includegraphics{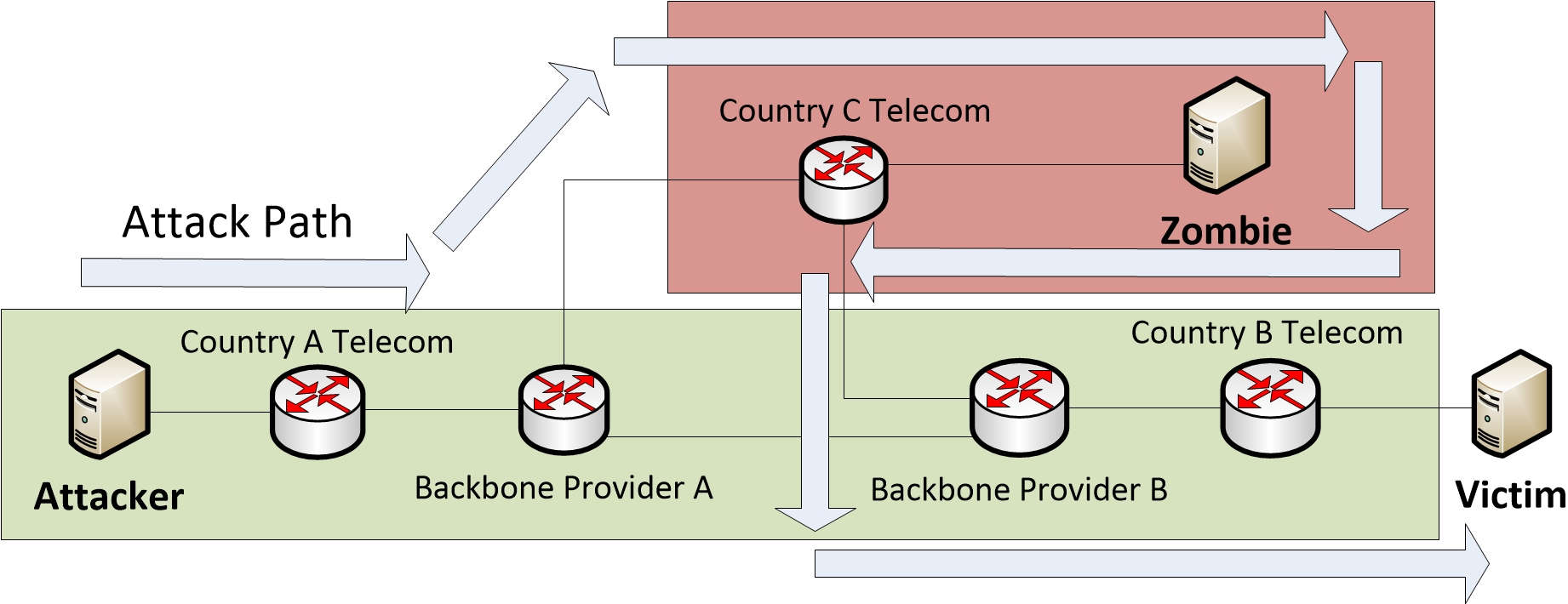}
\caption{The traceback problem: Country A and B are friendly, whilst Country C will not cooperate}
\label{fig:attribution1}
\end{figure*}

From the perspective of technical feasibility, a basic monitoring capability is feasible to establish using existing knowledge of cyber security practices, but significant challenges exist.  Firstly, caution must be used when the cyber terms are agreed.  Stating that country A will cease all cyber attacks upon country B is one that will be difficult to monitor.  Asking any organisation to monitor every network in a nation is unrealistic: it would simply require an amount of resources (both human and hardware) that would be infeasible to provide in the context of a peacekeeping operation.  It is therefore envisioned that peace agreements must list specific networks that should not be attacked.  The idea of a cyberspace safe layer as described by Akatyev and James~\cite{Akatyev2015} would fit this requirement, whereby a minimal set of critical systems that should be observed are identified.  Arguably the most significant technical obstacle is the attribution problem.  While it will be possible to observe an attack on a network, it will be difficult to prove where that attack originated.  Figure~\ref{fig:attribution1} shows how an attack can be routed through multiple countries and organisations, each with varying levels of cooperation and political relations.  This intentionally makes tracing the attack to its real source difficult.  Research into the attribution problem is ongoing~\cite{Nicholson2013,Wheeler2003,Robinson2014}, but as of today it will remain an obstacle towards fulfilling this OMR goal in cyber peacekeeping.  Statements from UN peacekeepers that a particular party has violated a peace agreement comes with significant implications towards peace and security in a region.  If solid, evidenced attribution cannot support such a statement without dispute, the feasibility of reaching this particular OMR goal is reduced.  The nature of cyber peacekeeping does present a potentially novel solution however.  As Wheeler and Larsen~\cite{Wheeler2003} noted, prepositioning of trust is an essential part of solving the attribution problem.  Cyber peacekeeping has the potential to enable this prepositioning in the form of what we call cyber peacekeeper reservists.  These are people who work at backbone providers and telecoms organisations, who are able to activate when required and cooperate to perform tracebacks across organisational and political boundaries.  This potential should be explored as part of future work.  At this stage, it is concluded that while the value of monitoring for actions which violate peace agreements would be high, the feasibility of doing so in the cyber domain is currently low, primarily due to the attribution problem.

\subsection{Human rights abuses}
Human rights is an issue that has always been at the core of peacekeeping, and monitoring for abuses of people's human rights is therefore the second observational goal of OMR.  The value this brings to peace and security is clear: the universal declaration of human rights~\cite{UDHR} is an internationally recognised document which sets out the basic rights of all humans, respect of which are essential in fostering peace.  To determine if this value can be transferred into cyber peacekeeping, we must consider if human rights can be threatened or violated in the cyber domain.

The answer to this question forms a separate and broad research area, and is actively being explored by many authors.  For example, Klang and Murray note that it is difficult to define how human rights relate to the growing use of cyberspace~\cite{Klang2005} whilst others such as Kulesza and Balleste~\cite{Kulesza2015} have focused on cyberspace's impact on the human right to privacy.

\begin{figure}[ht]
\centering
\includegraphics{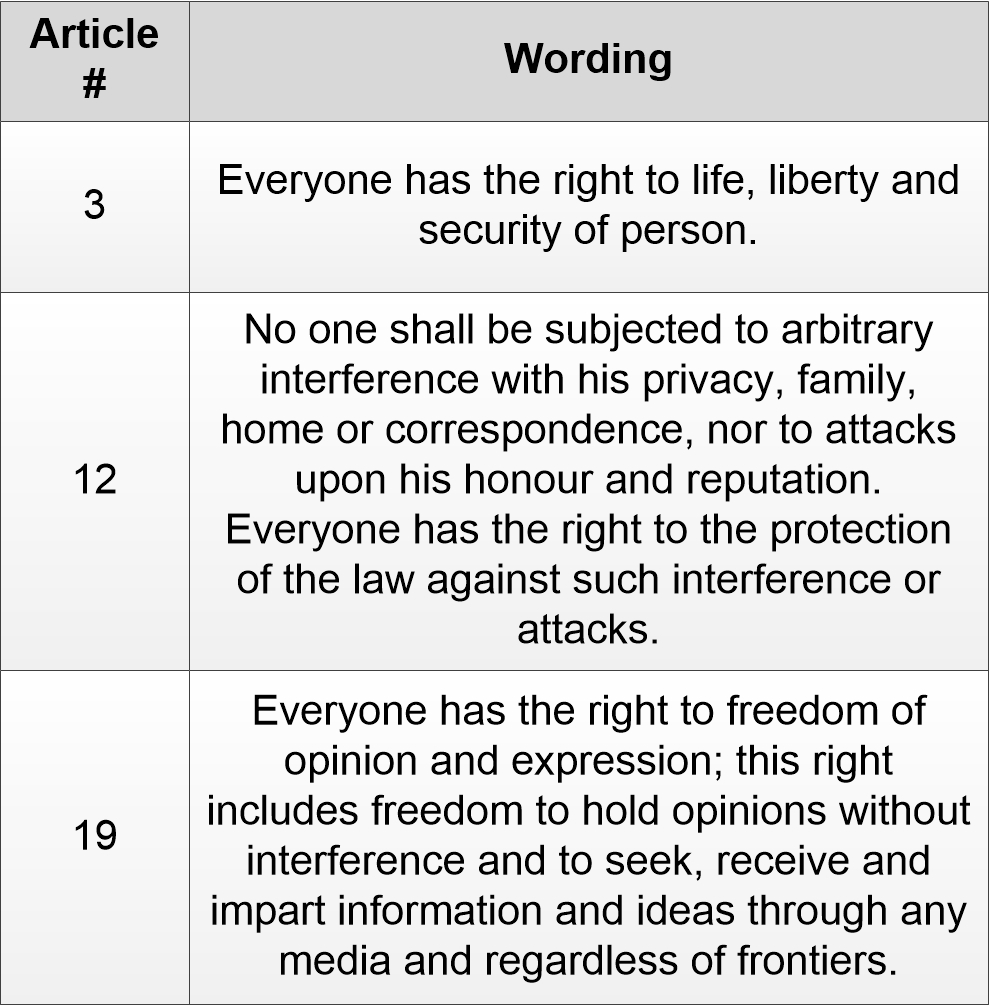}
\caption{Universal human rights at risk from the cyber domain}
\label{fig:HumanRights}
\end{figure}

Human rights specifically in relation to cyber warfare is less well studied.  Upon inspection of the rights laid out in the Universal Declaration of Human Rights (UDHR), we propose that three are particularly vulnerable to cyber warfare.  These are shown in figure~\ref{fig:HumanRights}.

Observing for threats towards life, liberty and security of person (Article 3) would be highly valuable in relation to cyber warfare.  Cyber attacks upon critical infrastructure such as public water supplies, the power grid or air traffic control have potential to threaten civilian life and security.  This would also be feasible to perform, since it does not rely upon attribution (cyber peacekeepers need only detect a threat to civilian life and provide warning, not attribute it to a particular party).

Observing for violations of privacy (Article 12) could also be threatened by cyber warfare.  Cyber weapons may expose the personal data or private communications of civilians and detecting such violations would be valuable towards maintaining peace.  However, there are some feasibility challenges here.  Snooping upon network traffic is fundamentally a passive activity which is difficult to detect technically.  For example, a government may inspect all traffic passing through a government controlled ISP, or compel regional technology firms to provide access to private customer data such as emails.  These are examples of human rights violations that are difficult to detect with a technical solution.  The political feasibility of observing for article 12 violations will also present a challenge.  We must remember that one of the core UN peacekeeping principles is that of consent.  This consent may be jeopardised if the host nation suspects that cyber peacekeepers are seeking to uncover government led privacy violations.  In this scenario, cyber peacekeepers may find it preferable to not observe for privacy violations in favour of maintaining the operation and hence the ability to detect more serious human rights violations such as article 3.

Monitoring for violations of freedom of expression and access to information (Article 19) would also be valuable towards maintaining peace.  If civilians can access multiple international news sources, they are less likely to be influenced by local propaganda and misinformation.  Furthermore, the ability to impart information regardless of frontiers may allow the international community to be alerted to violations happening locally.  A scenario can be envisioned whereby a nation blocks citizen access to certain information or denies internet access to people with certain views.  This is an area where the UN has already expressed an interest.  In 2016 UN special rapporteur David Kaye criticised Turkey's ``vast wave of internet shutdowns and content takedowns'' in the wake of the attempted July coup~\cite{Kaye2016}.  

Technically, if a cyber peacekeeping unit can secure access into a national internet service provider (ISP), there is potential to observe and report upon content blocking and other restrictions to access.  However, political feasibility may again present an obstacle due to the principle of consent.  Cyber peacekeepers must consider if consent for their operation would be maintained, if it is suspected that the cyber peacekeepers would be aiming to criticise the activities of the consenting government.  Political feasibility is therefore questionable.

In addition to the rights set out in the UDHR, there are efforts to develop a set of cyberspace human rights~\cite{IRPC2018} which aim to secure a new set of rights such as a right to access.  Should the world adopt such a set of rights, it is arguable that cyber peacekeepers would have a mandate to protect them.  Until such rights are agreed upon, it is difficult to predict their value and feasibility.

In summary, the value of monitoring for violations of human rights can be applied to a post-cyber warfare context.  We propose that the highest value will come from observing for threats to the right to life (article 3).  The feasibility of observing for violations will vary.  Monitoring for privacy violations is expected to be difficult, both for technical and political reasons.  Monitoring for violations of the right to seek and impart information will be similarly challenging.  Monitoring for threats to life will be more feasible, providing the necessary consent from the host nation can be secured.

\subsection{Changes in terrain, dispositions and civilian activity}
The final observational goal of OMR is to monitor and report upon changes in terrain, dispositions of forces and civilian activity.  This brings value towards the peacekeeping operation since it maintains situational awareness of the region.  Jeannot, Kelly and Thompson define situational awareness as "knowing what is going on around you''~\cite{Jeannot2003}.  A more detailed definition was given by Dominguez et al, who stated that situational awareness involved four specific steps:~\cite{Dominguez1994}

\begin{enumerate}
	\item Extracting information from the environment;
	\item Integrating this information with relevant internal knowledge to create a mental picture of the current situation;
	\item Using this picture to direct further perceptual exploration in a continual perceptual cycle; and
	\item Anticipating future events.
\end{enumerate}

Linking this back to peacekeeping, situational awareness helps to inform all aspects of the operation e.g. how patrols should geographically distributed, and highlight potential hot spots for future conflict.  In this regard, the value is really one that is indirect towards maintaining peace:  it enables peacekeepers to operate more effectively, enhancing the value of other activities they perform.  This "indirect" value can be applied to a cyber warfare context, but some translation effort is necessary.

To begin, observing for changes in `terrain' does not make sense in the cyber domain, since it is a domain in which terrain does not exist.  It is therefore proposed that rather than monitor for changes in terrain, cyber peacekeepers should monitor for changes in network structure.  As examples, the sudden unavailability of servers or the addition of new devices would be valuable changes to look for.  It must be noted however, that networks are naturally subject to change in normal conditions:  routing tables can change depending on network conditions and servers can become unavailable for patching and maintenance.  It will therefore be essential that cyber peacekeepers build up an understanding of what is normal and what is abnormal, effectively conducting anomaly detection upon the network they are monitoring.  Just as with monitoring for changes in terrain, the aim of monitoring for changes in network structure is to improve situational awareness.

Observing for changes in network structure is feasible, since extensive literature already exists on how this can be achieved.  NIST~\cite{NIST800-115} provides one such example, presenting guidance regarding network discovery, port and service identification, vulnerability scanning and wireless scanning.  By conducting activities such as these, cyber peacekeepers can gain a view of the cyber `terrain', observe for changes and hence maintain situational awareness of their environment.

Observing for changes in dispositions allows peacekeepers to monitor the location and make-up of military forces.  A rise or fall in troop numbers, or the movement of forces from one area to another are examples of changes in dispositions.  This again brings value through raising situational awareness.  As with the monitoring of network structure, being able to monitor dispositions of forces in cyber warfare would be valuable.

The feasibility of dispositions monitoring in the cyber domain is questionable.  The first obstacle is that observing cyber weapons and cyber combatants is challenging.  Cyber weapons are still not fully understood or defined~\cite{Robinson2014} and cyber combatants do not have to physically relocate or group up in order to launch effective attacks.  The question of how to monitor dispositions during cyber warfare could be an entirely new research area.  For the purposes of this article, a simple approach to this question is to consider what kind of situational awareness we hope to gain.  We propose the following two goals:

\begin{itemize}
  \item Identifying potential flashpoints of cyber conflict - what targets could a party be aiming to attack via cyber means?
	\item Identifying growth or decline in cyber warfare capability - what capability does each party hold, are they becoming more advanced?
\end{itemize}

Potential flashpoints of cyber conflict could be identified by observing for changes in cyber security dispositions.  For example, if cyber peacekeepers working in a nation witness a government's cyber security resources switching their attention towards protecting its power grid, this is arguably a change in dispositions which may indicate power systems as a point of cyber conflict in the near future.  The feasibility of this monitoring is questionable however.  While the movement of kinetic troops and vehicles can be observed, cyber troops can perform their duties at a variety of sites without physically relocating.

To identify a growth or decline in cyber warfare capability, cyber peacekeepers operating in a network could monitor and report upon the sophistication of cyber security techniques employed by local staff.  For example, the arrival of new cyber staff who are clearly more adept or observing more advanced attacks upon the infrastructure they are monitoring.  This would be feasible to perform, but we must again raise the challenge of attribution.  Even if an advanced attack was detected, the attribution problem would make any conclusion regarding the cyber warfare capability of any particular group difficult.

Finally, the observation of civilian activity brings value as an indicator of the level of peace and security felt by the local population.  Any change in the usual patterns of civilian life may point towards a change in the local situation that is worth investigating further.  It is proposed that in the cyber domain, this observational goal can be translated into changes in network traffic.  This would involve cyber peacekeepers building a baseline for normal traffic and subsequently looking for deviations from that baseline.  This is a feasible observation goal, since anomaly detection in network monitoring is an established field of research~\cite{Bhuyan2014,Manikopoulos2002} with a number of commercial products offering such a feature~\cite{Chakchai2016}.  It would also bring value to the peacekeeping process, by alerting to potential malware infections, denial of service attacks, data exfiltration and unauthorised access to peacekeeper protected networks.

\section{Cease-fire Supervision and Supporting Verification Mechanisms}\label{verification}
Cease-fire agreements are an important milestone towards ending a conflict and restoring peace.  They contain the terms agreed by each side, and can include any number of provisions. For example, the 2015 Minsk agreements in Ukraine~\cite{Minsk2015} set out twelve provisions that each side agreed to abide by such as the withdrawal of heavy weapons from certain areas, prisoner releases and early local elections.  There are three core goals of a cease-fire agreement~\cite{ceasefirebook2013}:

\begin{enumerate}
	\item Cessation of hostilities
	\item Separation of forces
	\item Monitoring and supervision
\end{enumerate}

The value that ceasefires bring is clear.  They deter parties from returning to conflict, make clear the rights and obligations of each group, create a sense of formal legal obligation and engage the international community~\cite{ceasefirebook2013}.  But it must be noted that they do not automatically lead to peace.  Forces at ground level may continue to fight, and an element of mistrust often exists between conflicting parties~\cite{Haysom2004}.  Peacekeepers help to maintain a ceasefire by acting as a trusted third party and verifying that each party is abiding by the agreed terms.

If future conflicts involve cyber warfare, it follows that future ceasefire agreements will contain cyber related terms.  There has not yet been an example of a ceasefire agreement which contains cyber terms, but using guidance from existing documentation~\cite{ceasefirebook2013,Haysom2004} along with current knowledge of cyber warfare~\cite{Robinson2014} it is possible to theorise about examples:

\begin{itemize}
	\item The cessation of all cyber attacks.
	\item Cessation of cyber attacks upon a specific infrastructure.
	\item Agreement to cooperate on cybercrime/spoiler attacks.  E.g. Each nation will pursue and prosecute lone wolf/spoiler cyber attackers inside their borders.
	\item Declaration of information stolen during the conflict.
	\item Declaration of systems compromised and assistance with returning control to rightful owners.
	\item Declaration of known vulnerabilities in opposing party's networks.
	\item Dismantlement of botnets.
	\item Remote disabling of malware (if possible) or assistance in locating and removing malware.
\end{itemize}

The value of these example agreements towards maintaining peace would be high.  They would provide confidence building in the peace process and expedite each party's ability to bring infrastructure back online and resume provision of essential services for civilians.  This has the potential to avoid state collapse or human rights violations.  The feasibility of verifying compliance with them is debatable however.  Regarding the cessation of all cyber attacks for example, the attribution problem again makes it difficult to conclusively determine that a cyber attack came from a specific party.  It may not be possible to provide unequivocal proof that a particular attack was conducted by a cease-fire signatory.  In this regard, any ceasefire term which relies upon attribution has low feasibility in practice.  This will be a major obstacle towards cyber warfare ceasefire agreements.

If attribution is an obstacle, terms which do not rely upon attribution are desirable.  The dismantling of botnets which are geographically located inside the cease-fire signatories country is an example of this.  This would be an activity that shows a commitment to reducing the potential for future cyber attacks originating from their region.  It does not require attribution; regardless of who is using the botnet, the action of dismantling it removes it from use by anyone.  This would be valuable towards peace and security, since it would reduce the opportunity for use in future conflict.

Regarding declarations of compromised systems and information stolen, cyber peacekeepers can work with both sides to determine if such declarations appear to be true.  This will involve an element of cyber forensics, in order to determine whether the declarations made align with the available forensic evidence on breached networks.  This kind of activity highlights the supervision aspect: both sides may be reluctant to discuss and share information about cyber issues with each other, since this information may be used against them should the conflict restart.  By disclosing the information to cyber peacekeepers, the information is kept with a secure and trusted third party for the purposes of restoring peace only.  It must be noted however, that by declaring to cyber peacekeepers the attacks used and systems compromised, there is a risk that a party would be giving up an advantage held over their opponent since the cyber peacekeepers may subsequently close vulnerabilities and remove any unauthorised control of systems.  In this regard there is a strong incentive to withhold information, in case the conflict reignites.  This incentive to cheat is counter balanced against the potential for being revealed as a cheat by cyber peacekeepers.  For example, if a nation does not declare that it has control of a nation's water supply and such control is discovered later, the nation will be revealed as trying to cheat the peacekeeping process.  

In summary, ceasefires containing cyber terms will become part of our future, but we have identified significant obstacles towards finding valuable and feasible ways of implementing them.  This challenge must be explored further in future work.

\section{Interposition as a buffer zone}\label{BZ}
Interposition as a buffer zone (BZ) is the final "traditional" UN peacekeeping activity.  The previous activity of OMR was a passive one: peacekeepers observe and report violations but do not interfere or become involved (except in the case of clear human rights violations, as highlighted by the Brahimi Report~\cite{Brahimi2003}).  Interposition as a buffer zone places peacekeepers into a more active role, whereby they not only observe and report but also act to prevent and stop violations.  The UN defines a buffer zone as ``an area established between belligerents and civilians that is protected and monitored by battalion peacekeeping forces and where disputing or belligerent forces and attacks on each other and the civilian population have been excluded''~\cite{UNIBAM1}.  This definition highlights the two key areas where a BZ brings value:

\begin{enumerate}
	\item Attacks on each other are excluded.
	\item Attacks against civilians are excluded.
\end{enumerate}

In a post cyber warfare context, we propose that this value can be preserved:  the field of cyber security already has extensive guidance and tools for preventing and neutralising cyber attacks on a network.  If a traditional BZ is a geographical area protected and monitored by peacekeeping forces, a CBZ would be an area of cyber space protected and monitored by peacekeeping forces.  Practically, this "area of cyber space" would be a specific network or site under the protection of cyber peacekeeping forces.  This can be formalised in the following definition:

\begin{definition}{Cyber Buffer Zone.}
A network or site that is protected and monitored by peacekeeping forces, where cyber attacks have been excluded.
\end{definition}

The significant question is how existing cyber security knowledge can be applied to a peacekeeping context to bring about the value of excluding cyber attacks.  To assist with this, we can examine how a traditional buffer zone operates.  

To exclude attacks, a traditional BZ is designed to ``maintain a visible presence and dominate the BZ with robust force projection to preserve the sanctity of the buffer zone by preventing any violation of ceasefire/peace agreement clauses''~\cite{UNIBAM2}.  From this, we can extract two distinct ways in which a traditional BZ excludes attacks:

\begin{enumerate}
	\item \textbf{Creating a visible and dominant presence}: The creation of a visible presence acts as a first line of defence through deterrence.  If an area is visibly well protected by peacekeeping forces, it acts as a deterrent to attackers since the risk of being observed and intercepted are high.
	\item \textbf{Robust force projection}: When deterrence fails, belligerents may be tempted to commit a violation inside the buffer zone.  Peacekeepers inside the zone act robustly to intercept violators and stop their actions with force if necessary.
\end{enumerate}

\subsection{Deterrence}
When considering the feasibility of a CBZ, it must be asked whether a visible presence and the associated effect of deterrence can be replicated.  This is important, since the deterrence effect of a BZ forms part of the value that it brings to maintaining peace and security.  A significant point to note here, is that the issue of deterrence in cyber warfare is a well researched area~\cite{Goodman2010,iasiello2014}, and there is ongoing debate regarding its effectiveness and challenges.  In a peacekeeping context, it is proposed that a CBZ could feasibly present a deterrent effect in the following ways:

\begin{enumerate}
	\item Raising awareness through announcements that a particular site is now under cyber peacekeeping protection - any attacks will be scrutinised and the risk of detection increased.
	\item Threat of enhanced trace-back - Using the concept of cyber peacekeeper reservists based at backbone providers, there is an increased likelihood that an attacker will be identified and held accountable.
	\item High risk of attack failure - A site protected by a CBZ will be regarded as a hard target for attack, this will have a deterrent effect since the risks (loss of a cyber weapon, revealed as a ceasefire violator) outweigh the reward.
\end{enumerate}

While not a visible presence in the traditional sense, these aspects would arguably produce the same effect: to influence the thought process of a potential attacker so that they decide not to launch an attack.  An alternative view is that they also hold potential to have the opposite effect.  For example, a hacking group may see a UN protected network as an enjoyable challenge.  In this regard, a deterrent effect is something that the UN must fight for and win at an early stage.  UN cyber peacekeeping must demonstrate that it can use the cyber reservist scheme to successfully trace back attacks and identify attackers, regardless of their geographical location.  Once this effectiveness is proven through high profile successes, a deterrent effect will be established for future operations.  If a cyber peacekeeping unit fails to do so, the deterrence effect is reduced and future CBZs will be less effective.

\subsection{Robust force projection}
In the context of a CBZ, robust force projection will be active measures taken by cyber peacekeepers to detect and neutralise cyber attacks in the network they are protecting.  This is largely the application of existing cyber security knowledge.  Cyber peacekeepers would work to secure the network both in the short and long term.  The ultimate aim would be to strengthen cyber security and local skills to a level where cyber peacekeepers could withdraw.  A possible list of activities are as follows:

\begin{itemize}
	\item Dropping packets suspected of being cyber attacks.
	\item Blocking attacker IP ranges at network devices.
	\item Performing host hardening (patching, removing unnecessary services).
	\item Providing additional capacity to reduce impact of DoS attacks.
	\item Providing training for local staff to enable eventual peacekeeper withdrawal.
\end{itemize}

All of these activities should be feasible to perform, with the aim of detecting and neutralising cyber attacks before they can damage their intended target.  A precise list of activities can be explored as part of future work, and will likely vary depending on the network to be protected.  For example:  a CBZ at a nuclear power plant will implement different methods of defence in comparison to a CBZ at a governmental office.

It is concluded that a cyber buffer zone would bring value to the peacekeeping process by excluding cyber attacks upon networks which are critical towards maintaining peace and security.  It is also a feasible activity, since it utilises existing cyber security knowledge and tools to provide the buffer.

\section{Disarmament, Demobilisation and Re-integration (DDR)}\label{DDRsection}
The UN defines DDR as an activity which ``aims to deal with the post-conflict security problem that arises when combatants are left without livelihoods and support networks''~\cite{UNDDRfield2014}.  It is a process which identifies ex-combatants and assists them in finding peaceful and sustainable civilian life.  This is an important activity, since those who have known nothing but war may work against peace if they cannot sustain themselves in a peaceful environment.  Each of the three sub-activities - disarmament, demobilisation and reintegration will be examined to determine their potential value and feasibility in a cyber context.

\subsubsection{Disarmament}
\begin{figure}[ht]
\centering
\includegraphics{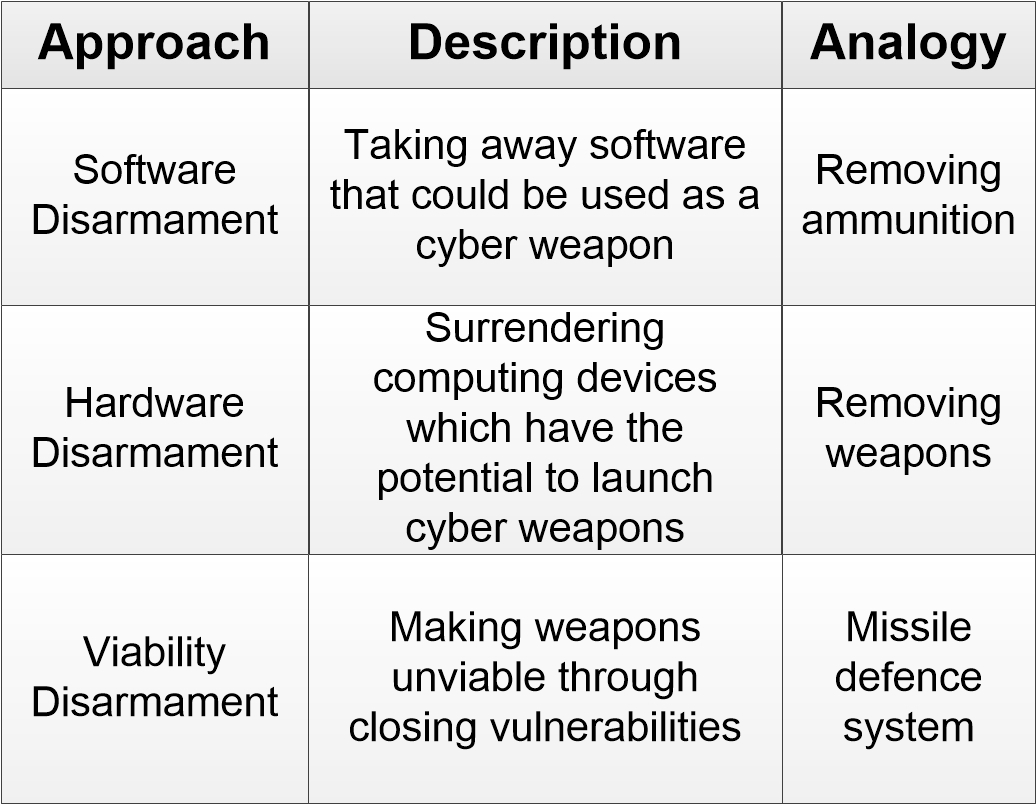}
\caption{Three potential approaches to cyber disarmament}
\label{fig:disarmament}
\end{figure}

Disarming ex-combatants during peacekeeping brings values in three ways.  Firstly it is a symbolic act, the giving up of weapons reinforcing the message to the ex-combatant that the conflict is over.  Secondly, it serves to physically remove weapons from the region which reduces the likelihood of a relapse into armed conflict.  The final purpose is to act as a confidence building measure.  When civilians and ex-combatants from both sides witness weapons being surrendered, it provides reassurance that the peace process is real and that it is progressing.  When considering if this value could be maintained in a post-cyber warfare context, three potential approaches can be identified, summarised in figure~\ref{fig:disarmament}.

The first approach is software disarmament, which aims to remove software which could be used as a cyber weapon such as hacking tools.  This approach has a number of feasibility issues however.  Firstly, there is no widely accepted definition of a cyber weapon~\cite{Arimatsu2012,Robinson2014}.  Secondly, software can have both peaceful and warfare uses~\cite{Robinson2014}, which makes choosing which software to confiscate difficult.  Thirdly, cyber weapons have characteristics which kinetic weapons do not.  Fir example, ease of replication allows cyber weapons to be acquired, passed on and proliferated at very little cost.  If confiscated software can simply be re-downloaded, it is never truly removed.  Software can also be easily hidden through the use of encryption and remote storage (e.g. to cloud service).  These problems cause software disarmament to fail at the intended goals: it will not reduce the number of weapons, and subsequently not increase confidence because each party will be aware of its weaknesses.  Whether it can act as a symbolic gesture is debatable.  A cyber ex-combatant may find some psychological benefit from surrendering hacking tools to peacekeeping forces and is an aspect worth exploring in future work.  Despite the problems with software disarmament, there are instances where it could be argued as having merit.  For example, if a party developed a cyber weapon using zero-day exploits and kept strict controls over its replication, a software disarmament process could arguably destroy all copies and leave none behind.  This argument has weaknesses however, since the vulnerability (and knowledge of it) would still exist, allowing the weapon to be recreated in the future.

The second approach is hardware disarmament, whereby computing devices such as laptops, PCs and tablets are surrendered.  Without access to such devices, ex-combatants will not be able to resume launching cyber attacks.  A significant feasibility problem with hardware disarmament is that the confiscated hardware, like software, is dual use.  A laptop used for launching cyber attacks can also be used for social, economic and educational purposes.  Taking this hardware away from an ex-combatant may actually harm the peace process by removing devices that are needed to flourish in a peaceful society.  A further problem is that in wealthy nations, removing hardware is barely more than a temporary inconvenience.  It must be concluded that this approach will not fulfil the goals of disarmament.

A third approach is viability disarmament.  Here the focus is shifted away from the attacker and towards the target.  By hardening targets and closing vulnerabilities, cyber weapons can be ``disarmed'' remotely.  This is an appealing approach, since it avoids the problem of having to locate every copy of a cyber weapon.  This method exploits the fast life cycle characteristic of cyber weapons~\cite{Robinson2014}.  Considering the disarmament goals, this approach comes out favourably.  It would not technically reduce the number of cyber weapons in a region, but it would reduce the effectiveness of those weapons on targets that are essential to peace and security.  It would also be a confidence building measure, since both parties would understand that making cyber weapons ineffective is a technical solution that works.  Regarding symbolism, viability disarmament requires no voluntary surrendering of weapons and therefore this goal is missed.

In summary, no single approach to cyber disarmament appears to fulfil all of the traditional goals.  Viability disarmament provides confidence building and contributes towards reducing the number of effective weapons in the region.  But this approach lacks the symbolism of an ex-combatant voluntarily surrendering their weapons.  To achieve all of the goals of disarmament, a combined approach is necessary.  Viability disarmament alongside an encouragement to voluntarily surrender software and tools is arguably the approach that would fulfil all of the disarmament goals.  This would be a feasible approach, and also one that would bring value to the peacekeeping process.

\subsubsection{Demobilisation}
Demobilisation is the physical and mental process of taking a combatant and supporting them in a transition back to civilian life.  It is defined by the UN as ``the formal and controlled discharge of active combatants from armed forces or other armed groups''~\cite{UNDDRfield2014}.  There are two overall goals of demobilisation.  The first is the physical goal of separating the combatant from their command and control structure.  This is followed by the mental goal of changing their mindset from that of a combatant to that of a civilian.  Demobilisation is attractive to ex-combatants due to the assistance they receive. In the short term this assistance includes food, shelter, training, education and tools.  The value of performing demobilisation is that combatants are guided through a process which helps to ensure that they have peaceful ways of living and are not reliant on conflict to maintain a livelihood.  When considering if this value could be maintained in cyber peacekeeping, it is necessary to ask if cyber combatants require demobilisation.  This is not a simple question to answer, since there are generally two types of combatants recognised by the DDR process - those from armed forces (national militaries) and those from armed groups (informal factions).

For both category of cyber ex combatant, it is proposed that demobilisation would be valuable.  Cyber combatants will have a skillset which has the potential to damage the peace process.  For example, without a demobilisation process they may turn to cyber crime in order to sustain themselves during peacetime.  Demobilising them can provide education on opportunities that leverage their highly desirable skills for a peaceful and sustainable purpose (e.g. cyber security roles).

In the case of cyber combatants from armed forces, this is feasible to perform, since the cyber combatants will be referred to the DDR process by their commander.  Cyber combatants from armed groups present more of a feasibility challenge.  Membership of armed groups may be informal, without any written documentation or proof of membership, making identification of members difficult.  Armed groups may also simply disband and mix in with civilian communities once hostilities have ended.  The benefits that DDR offers are designed to overcome this by encouraging self-identification~\cite{UNDDRfield2014}.  To ensure that DDR benefits are not being claimed fraudulently, self identifiers are tested for knowledge such as key battles and familiarity with weapons.  Such tests could be adapted for cyber by testing for detailed knowledge of key cyber attacks: which vulnerabilities were used, which systems were breached and so on.

Another feasibility issue with armed groups is that the nature of cyber warfare allows for cyber combatants to be physically located outside the conflict region.  This problem is unique to cyber peacekeeping, since kinetic combatants must generally be physically present in a region to conduct attacks.  The closest guidance on this issue can be found in the UN DDR Framework Module 5.40~\cite{UNDDRframework2014}.  This module addresses the problem of foreign combatants crossing borders to conduct warfare in other nations.  The suggested solution is to intern these combatants and return them to their country of origin without offering any DDR services. If this guidance is maintained into cyber peacekeeping, DDR should not be offered to foreign cyber combatants.  Another potential avenue is to simply regard these people as outside spoilers:  those not involved in the peace process, e.g. disgruntled groups, citizens and combatants~\cite{Stedman1997}.  If they are regarded as spoilers rather than ex combatants, demobilisation is not provided.  A counter argument can be made however, in that providing DDR would bring value if it meant that the cyber attacks ceased and future cyber attacks from this person were avoided.  Ultimately, the feasibility of tracking down such people and the pressures on funding are likely to result in this value being expensive to realise.

In summary, demobilising cyber combatants would bring value to a peace operation since it would guide cyber ex-combatants away from cyber crime during peacetime.  It is also feasible to perform, since existing procedures can generally be followed with some adaptations.  Future work should explore how this activity could be developed even further.

\subsubsection{Reintegration}
Demobilisation is a short term process, and is followed by the longer term process of reintegration.  It is formally defined as ``the process by which ex-combatants acquire civilian status and gain sustainable employment and income''~\cite{UNDDRfield2014}.  The value of reintegration as an activity is clear.  It provides life skills, vocational training, education and on the job training to ex-combatants who have potentially little experience of living outside of conflict and war.

When considering if reintegration would bring value for cyber ex-combatants, a number of points must be considered.  Firstly a cyber ex-combatant may already be integrated with society, holding employment that they can return to.  As an example, a cyber security professional may have been conscripted to fight for the government and can now return to their previous employment.  In this regard, the reintegration process will either be not required or be very brief.  Others may be on the opposite end of the scale and truly lack knowledge on how their skills could be applied to civilian life.  An example here is a teenage cyber combatant who has never held legal employment and would struggle to transfer their skills without some support.  Reintegration would be valuable here.

Performing reintegration with cyber ex-combatants is feasible.  Programmes can be developed that help ex-combatants capitalise on existing cyber skills and help to guide them into a sustainable livelihood.  In particular, there are natural links that exist between the DDR process and the Security Sector Reform (SSR) process~\cite{Mcfate2010}.  As national police and security sectors are reformed, ex-cyber combatants are in a position to fill roles aimed at bolstering the cyber defence of the nation and to receive on the job training and development.  A model such as the Estonian Defence League is a good example of what is possible~\cite{Kaska2013}.  Private industry also has an incentive to actively take part in such reintegration efforts, since they will directly benefit from access to highly desired cyber security skills.  In conclusion, reintegration is expected to be both valuable and feasible.

\section{Security Sector Reform (SSR)}\label{SSR}
Security Sector Reform (SSR) aims to leave a nation with a capable and suitable security sector.  The UN gives SSR the following definition:  ``a process of assessment, review and implementation as well as monitoring and evaluation led by national authorities that has as its goal the enhancement of effective and accountable security for the State and its peoples without discrimination and with full respect for human rights and the rule of law''~\cite{UNSSR2012}.  A nation's security sector includes defence, law enforcement, corrections, intelligence services, border management, customs and judicial systems.  The types of reform these institutions require after a conflict can vary, ranging from sweeping structural reforms to small procedural changes.  In all cases, the aim is to enhance the effectiveness, efficiency, accountability and affordability of the institution~\cite{UNDefenceSector2011}.  The value of SSR is clear:  a well performing security sector can uphold law and order in a nation, preventing the re-emergence of conflict.

When considering a post-cyber warfare context, SSR will bring value towards restoring peace.  The institutions of national defence, law enforcement and intelligence services will have an interest in developing a strong cyber capability in order to tackle threats from cyberspace.  Developing this capability is also in the interests of the UN, since it contributes towards maintaining long term peace and security in the region and will enable the eventual withdrawal of cyber peacekeeping forces.

How this cyber capability could be built, and how it should look are important questions.  It is proposed that the development of a cyber capability can be assimilated into already existing procedures.  For example, if the cyber security capability of a particular institution is too strong and is being oppressive, it may require reigning in.  On the other hand, it should be capable enough to detect and counter cyber crime and foreign cyber attacks.  However, it must be noted that finding this balance will be politically challenging, as each nation may have its own opinion of what level of capability is reasonable.   There are also clear links between the DDR and SSR processes.  The DDR process aims to find legal, peaceful employment for ex-combatants, while the SSR process requires human resources with certain skills.  Demobilised cyber ex-combatants are therefore an important resource for the SSR process.  If the DDR process can harness their existing cyber abilities, there is potential to find long term employment in a reformed national cyber defence sector.

UN SSR guidance~\cite{UNSSR2012} states that any reform shall not include activities which compromise the sovereignty and territorial integrity of other states, human rights, or activities contributing to internal conflicts in the host nation.   The UN therefore provides basic training only, and the programme is carefully designed to not provide any support that would enhance a country's capability to wage war or violate human rights~\cite{UNDefenceSector2011}.  The same approach must be taken towards developing a cyber capability.  Training should be limited in scope to performing cyber defence, and policy reforms should bolster respect for human rights such as privacy and the right to seek and impart information.  In summary, SSR in the context of cyber will not only be valuable, but essential towards ensuring long term peace.  It is also expected to be feasible to perform, but future work should aim to identify any unexpected obstacles.

\section{Electoral Assistance}
UN electoral assistance as an activity has two primary aims.  Firstly, to assist a nation in holding credible and legitimate elections in line with internationally recognised standards.  Secondly, to build a national capacity to hold future elections without assistance~\cite{UNElections2007}.  The value of this is twofold.  Firstly, periodic and genuine elections are a human right set out in the Universal Declaration of Human Rights~\cite{UDHR}.  Second, it brings confidence to the process and to the result:  even if your chosen candidate does not win, the process was fair and there is no justification for resorting to violence.

The act of voting and the act of counting votes is largely a kinetic one, even in the most developed nations.  In an environment of physical voting booths and ballot papers, the cyber domain has arguably little relevance to electoral assistance.  There are however some instances where a cyber capability could be valuable.

\subsection{Election Hacking}
The issue of ``election hacking'' was brought into prominence after the 2016 US presidential election, with US intelligence agencies claiming that Russia intentionally conducted hacking to influence the election outcome in favour of Republican candidate Donald Trump~\cite{Shane2017}.  While the accusations do not suggest that any voting machines or counting processes were directly tampered with, it is claimed that Russia hacked and leaked information to influence the result.  While the election process was not directly hacked, the question of if it is a ``fair'' election in the face of cyber attacks designed to influence the voters is debatable.  Similar questions have been raised over the French election in 2017~\cite{Willsher2017} and even the Brexit referendum in the United Kingdom~\cite{Adam2017}.

The USA, France and UK are politically stable enough to experience such electoral doubt without experiencing a threat to internal peace and security.  In post conflict environments, peace will be more fragile with the smallest of doubt over an election result potentially leading to violence.  Cyber peacekeepers could therefore bring value in combating both hard (e.g. counting process, voting machines) and soft (misinformation, fake news) cyber attacks on future elections.

With regards to hard cyber attacks on an election, it must be noted that even in developed nations, the majority of elections are paper based and not heavily reliant upon the cyber domain.  The Netherlands abandoned electronic voting in 2008 after such security concerns led to a loss of confidence by the public~\cite{Jacobs2009}.  In this regard, most elections are unlikely to be threatened by hard cyber attacks.

The feasibility of combating soft attacks on elections is more challenging.  The spread of fake news is a challenge the world is already trying to combat, with the UK setting up a government unit specifically for that task in 2018~\cite{Walker2018}.  The UN may find it necessary to open a similar unit inside of conflict regions where an election is due, robustly pointing out misinformation designed to influence the outcome and educating civilians on how to spot it themselves.  It is feasible that cyber peacekeepers could also provide assistance to political parties in the run up and during elections.  This assistance would include the securing of party computers and networks, with the aim of preventing cyber attacks designed to leak information regarding the candidates.  In effect, there is potential for cyber peacekeepers to perform OMR or a cyber buffer zone on political party networks in the run-up to an important election.

Value could also be found in securing parts of the election infrastructure that do have a cyber component.  For example, information about candidates and voters may be stored in a database.  Cyber peacekeepers may provide value here, by securing this database and preventing attacks upon its availability, integrity and confidentiality.  There is also the potential for cyber peacekeepers to provide training and prevention of social engineering attacks upon candidates and their staff.

In summary, the value that cyber peacekeepers could bring to electoral assistance largely depends upon the level of cyber dependence in the election.  In cases where the election is paper based, value will be minimal.  Should the UN adopt electronic voting and tallying, the value will rise in the form of protecting from hard cyber attacks.  Soft cyber attacks on elections are a relatively new concept, and there is potential value to be found in combating misinformation and fake news designed to influence the result.  The feasibility of such an activity should form a basis for future work.

\section{Mine Action}
Mine action is an activity designed to reduce the threat and impact of mines and explosive remnants of war (ERW), including cluster munitions, on peace and security, humanitarian relief, human rights, and socio-economic development~\cite{UNmineaction2012}.  It has four overall goals.  Firstly it aims to reduce risk to civilians through surveying, marking, fencing and clearance of affected areas.  Secondly mine action works to support victims of mines by working with a nation to secure access to rights and services.  A third goal is to develop a national capacity whereby the nation can continue to perform mine action in the long term without external assistance.  The fourth goal is to educate and promote the issue of mine action at a national and international level.

The value that mine action brings to the peace process is high.  It allows civilians to resume working the land, opens up routes to enable trade and prevents physical harm.  Translating the value that mine action brings into the cyber domain is challenging, since mines are a problem that exist in the kinetic world.  Although mine action cannot be directly translated, there are striking similarities between the land and sea problem of mines and the cyber problem of malware.  Just as a field can be littered with mines during conflict, a computer system can be littered with malware.  Both mines and malware remain hidden until activated or detected, and the harmful effects persist long after a conflict has ended.  These effects present threats to peace and security, especially if they are located in critical national infrastructure.  It is therefore worth investigating whether ``malware action'' could be a feasible activity which retains the value that mine action brings.

\subsection{Malware Action}
\begin{figure}[ht]
\centering
\includegraphics[width=\columnwidth]{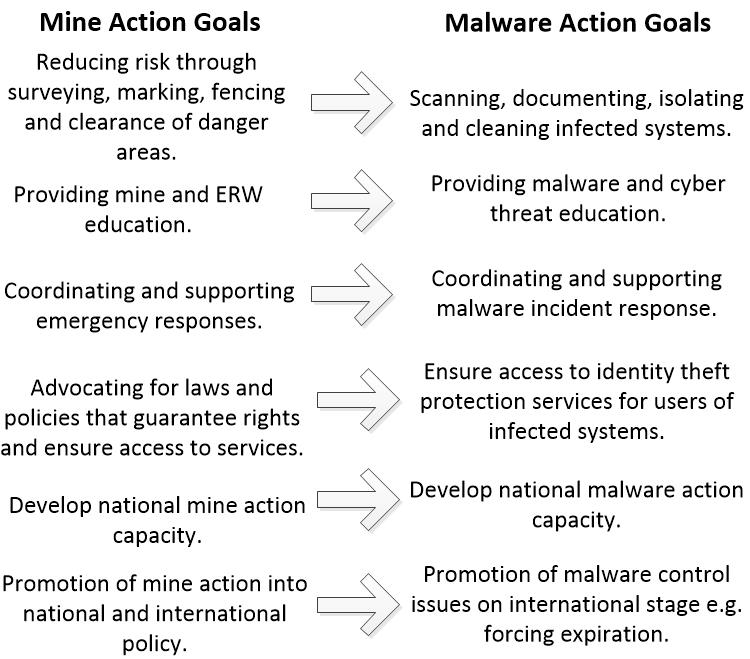}
\caption{A comparison of goals between mine and malware action}
\label{fig:malwareaction1}
\end{figure}

To explore whether malware action could bring value to a peacekeeping operation, figure~\ref{fig:malwareaction1} presents the goals of mine action and suggests equivalent goals of malware action.

Figure~\ref{fig:malwareaction1} shows that a cyber equivalent goal that closely matches can be found in all cases.  This suggests that there may be value in performing malware action and that further investigation is justified.  Malware action is also a feasible activity, since technical guidance on how to respond to a malware incident is well established in cyber security literature.  One such set of guidelines is provided by the National Institute of Standards and Technology (NIST)~\cite{NIST2005}.  Using these guidelines, it is possible to propose how malware action could be implemented.  A seven phase approach towards malware action is as follows:

\textbf{Phase one}: A request is made for malware action upon a set of devices or more generally at a particular site of importance.  This is an essential first step, and highlights that malware action only takes place with consent and invitation of the network owner.

\textbf{Phase two}: Following a request for malware action, a plan will be developed.  This will include a number of activities:

\begin{itemize}
  \item Identification of devices and systems
  \item Prioritisation of systems
	\item Assignment of resources
	\item Identification of partners
	\item Stakeholder meetings to ensure ownership at all levels and understanding of the process.
	\item Identification of goals, risks and responsibilities.
\end{itemize}

\textbf{Phase three}: Surveying of target devices begins.  Cyber peacekeepers examine systems for signs of malware infection.

\textbf{Phase four}: If a device is determined to be infected with malware, it is marked as such in documentation and cyber peacekeepers work to contain the malware to prevent it spreading and causing additional harm.  This is equivalent to the fencing activity in mine action.

\textbf{Phase five}: Cyber peacekeepers work to remove the detected malware.  This is equivalent to the clearance stage of mine action.  Technical guidance on malware clearance is given in section 4.4 of NIST SP800-83~\cite{NIST2005}. 

\textbf{Phase six}: Monitoring and evaluation. Cleaned devices are monitored over time to confirm that the malware has been fully removed.  The clearance process is evaluated for future improvements.

These first six phases match the activities of surveying, marking, fencing and clearance.  With regards to malware, a seventh phase is needed which is not included in existing mine action documentation.

\textbf{Phase seven}: Recovery of devices and systems to original state and removal of containment measures.  This phase is analogous to removing barriers around a road once it is cleared of mines, allowing it to used for its valuable peaceful purpose.

Other goals of malware action such as providing malware education and coordinating emergency malware response teams are also feasible, and will likely form a part of cyber peacekeeping malware action.

\section{Protection and Promotion of Human Rights}\label{HRsection}
Human rights are firmly established at the core of all peacekeeping operations, and are central to all of its activities.  For example, OMR observes for human rights abuses whilst electoral assistance aims to ensure everyone's human right to genuine and period elections.  But in addition to being part of every peacekeeping activity, the protection and promotion of human rights is also an activity in itself.

Firstly, human rights issues are highlighted and made explicit in training and planning documentation.  Points at which a peacekeeping activity has the potential to threaten human rights are made clear, and advice is given on how to proceed.  This helps to avoid cases where the activities of peacekeepers themselves have the potential to violate human rights.  Secondly the UN Office of the High Commissioner for Human Rights (OHCHR) supports all peacekeeping operations, providing expert advice regarding human rights on a day to day basis~\cite{OHCHR2014}.

\subsection{Protection}
Human rights can be threatened both from external actors and inadvertently by peacekeepers themselves.  For example, a buffer zone may infringe upon a person's right to freedom of movement within a country's borders.  By considering how each activity could threaten human rights, changes can be made to minimise that threat or manage it in a suitable way.

To consider the value of protecting human rights during cyber peacekeeping, it must be asked whether human rights have the potential to be violated both by cyber peacekeepers and by others in the cyber domain.  This was briefly discussed when making a justification for cyber peacekeeping, where it was argued that human rights can be threatened by cyber attacks.  In particular, the right to life and security of person was offered as a justification for the need of cyber peacekeeping.  In this section, a deeper exploration of the human rights issues that surround the concept of cyber peacekeeping is provided.

Research into human rights in cyberspace is extensive~\cite{Murray2004,brophy1999,citron2009,walters2002}, with authors highlighting sexual exploitation, freedom of speech, digital divides, censorship and privacy as areas where cyberspace presents human rights threats.  The UN Human Rights Council~\cite{HRC1727} has emphasised the ``unique and transformative nature of the Internet not only to enable individuals to exercise their right to freedom of opinion and expression, but also a range of other human rights, and to promote the progress of society as a whole''~\cite{HRC1727}.  Other examples of the UN associating human rights violations and cyber are found in the expression of concern at the right to privacy being regularly violated in cyber space~\cite{HRC2737} and the passing of a resolution in 2016 condemning any nation which intentionally disrupted the internet access of its citizens~\cite{UN2016HR}.  Looking at the evidence, there is no doubt that the UN regards the cyber domain as a place where human rights can be threatened, and that online violations of human rights should be taken as seriously as offline violations.

From the perspective of avoiding breaches by cyber peacekeepers, arguably the biggest threat to avoid is a violation of privacy.  Due to the nature of OMR and other activities such as malware action, there is potential that cyber peacekeepers will have access to personal information and communications.  Kleffner and Dinnis~\cite{Kleffner2013} provide some insight here.  They considered how cyber peacekeepers themselves might inadvertently threaten human rights, stating that ``interference with cyber infrastructure or data must be carried out in compliance with the requirements of human rights law''~\cite{Kleffner2013}.  However, they also note that rights such as privacy and freedom of expression are not absolute: international human rights law does allow certain interference where national security and public order issues are present~\cite{Kleffner2013}.  In this regard, there is a balancing act to be made between allowing cyber peacekeepers to effectively maintain peace whilst also being careful not to breach human rights unnecessarily.  This is therefore a risk that will need to be managed, and in line with existing procedures could be achieved via the following:

\begin{itemize}
  \item Interweaving human rights considerations into cyber peacekeeping policy, training and evaluation. This would provide warnings of where extra caution must be taken and practical measures to minimise breaches.
  \item Daily support from the OHCHR regarding privacy issues as they arise.  The OHCHR will require cyber-legal expertise to provide support to cyber peacekeepers in the field.
\end{itemize}

These are feasible activities to perform, since they already take place as part of the existing peacekeeping framework.  As each cyber peacekeeping activity is developed, human rights must be considered and guidance provided on how to ensure their protection.  This would clearly be a valuable activity, and one that should be performed.

\subsection{Promotion}
There are also opportunities for cyber peacekeeping to bring value by promoting human rights.  At a local level, cyber peacekeepers are in a position to influence states to adopt policies and legislation that respects human rights in cyberspace.  Arguably the best path to achieve this is through security sector reform (SSR).  During SSR, a respect for privacy and other human rights can be promoted and included into policy and legislation reforms.  This is a feasible activity to perform, since cyber peacekeepers are taking a leading role in the reform of national security institutions, a position which provides real opportunities for human rights promotion.  It is also prudent to consider the value of promoting cyber related human rights on the international stage as a standalone activity.  It is proposed that this activity would be more suited to other organisations and UN departments, and not to cyber peacekeeping.  This is because cyber peacekeeping will be deployed in response to a specific situation, and with a specific mandate.  Promoting cyber related human rights on a general level is unlikely to be part of this mandate, and more likely to fit better into work of other actors such as the UN Human Rights Council.  It must be noted that the promotion of human rights will likely be one area where political obstacles will be encountered:  not all nations will share the UN's perspective on human rights, and some will resist attempts to impose certain values upon their nation.

In summary, cyber peacekeeping must protect against violations of human rights from two sources: external parties and from cyber peacekeeping itself.  Protecting from external sources is largely covered in other activities such as OMR, electoral assistance and malware action.  It therefore requires no separate action as part of this activity.  To protect against violations from cyber peacekeeping itself, it has been proposed that the risk can be managed through an emphasis on human rights issues in cyber peacekeeping training and planning documentation.  The Office of the High Commissioner for Human Rights can also provide further support, although this office will require the necessary cyber-legal expertise to do so.

\section{Restore and Extend State Authority}\label{RESA}
Following a conflict, a nation's control over its territory and borders may be weak or absent.  Without the ability to operate a government and provide basic services to the population, the stability of the nation and security of its population is under threat. The restoration and extension of state authority (RESA) is therefore one of the activities of a multi-dimensional peacekeeping operation.  The UN Department of Civil Affairs lists the following tasks as part of RESA~\cite{UNCAH2012}:

\begin{itemize}
  \item \textbf{Supporting development of transparency and accountability}: State authority relies on gaining the trust of citizens. Encouraging the government to be transparent and accountable in its actions is essential.
	\item \textbf{Assessment of capacity and support needs}: A review of the current situation and what assistance is required to improve it.
	\item \textbf{Performance Monitoring}: Identifying where weaknesses lie and how they could be improved.
	\item \textbf{Civic Education}: Citizen participation in areas such as voting is crucial, and this task aims to educate citizens on their role in the process.
	\item \textbf{Logistic and Administrative support}: Providing basic logistical and administrative assistance.
	\item \textbf{Small-scale capacity-building support}: E.g. training elected officials in local finance or budgeting.
	\item \textbf{Supporting policy, planning and decision-making processes}: State institutions may be starting from scratch and will require assistance in developing the basic processes undertaken by a state.
	\item \textbf{Support to resource mobilization}: Directing donor resources to areas of most need.
\end{itemize}

The UN makes it clear that its role in these tasks is always a supporting one, rather than as a substitute for the state.  Terms such as enable and facilitate are used, to avoid creating a situation where the UN is seen to be providing services, rather than the state.  By encouraging the state to be the provider, trust between citizens and government begins to develop.  To explore whether cyber peacekeeping could bring value to the restoration and extension of state authority, it is necessary to ask whether a state's ability to generate revenue and provide basic services can depend upon cyberspace.  A problem here is that there is no definition of ``basic service'', since it can vary depending upon the culture and expectations of the population.  The UN Department of Civil Affairs~\cite{UNCAH2012} provides some guidance, proposing that water, sanitation, health and primary education are basic services common to all nations.  Taking this list, it must be argued that the value of cyber peacekeeping towards providing these services is largely dependent upon a nation's level of cyber dependence. Two scenarios highlight this point.

\subsection{Scenario A: Low Cyber Dependence}
Nation `A' has a low level of cyber dependence.  The government revenue office works on standalone computers with local databases.  The water supply is delivered from a central plant which has no external cyber connections.  The value that cyber peacekeeping could bring in restoring state authority here is small.  Restoring the functioning of the water plant or resuming operation of the tax office will likely be kinetic tasks with no need for cyber capability.

\subsection{Scenario B: High Cyber Dependence}
Nation `B' has a more cyber dependent infrastructure.  Government departments are interconnected via an integrated IT system, which allows the sharing of information between departments, cloud storage and online access to government services for citizens.  Power is supplied by a smart grid whilst driverless trains and smart traffic systems connect cities.  Cyber peacekeeping will clearly bring greater value to the peace process here since restoring the functioning of these cyber dependent systems and fending off future attacks will be crucial in restoring basic services to civilians.

\begin{figure}[ht]
\centering
\includegraphics[width=\columnwidth]{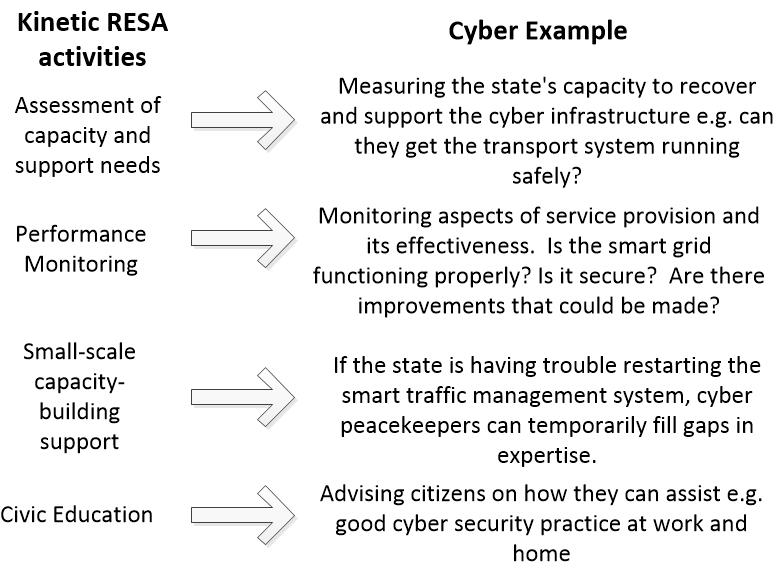}
\caption{Examples of restoration of state authority activities in cyber peacekeeping}
\label{fig:resa-translations}
\end{figure}

In cases where cyber peacekeeping would prove valuable, it is also expected to be feasible.  Evidence for this comes from the fact that it is possible to present examples of how existing activities could be applied to the cyber domain.  Some examples to demonstrate this are provided in figure~\ref{fig:resa-translations}. 

In summary, cyber peacekeeping will bring value to the restoration and extension of state authority only in cases where state cyber dependence is moderate or high.  In the aftermath of cyber warfare, states may find difficulty in bringing cyber dependent infrastructure back online, and this has the potential to weaken state authority and subsequently delay the return of peace and security in the region.

\section{Supporting Role Activities}
The final two activities of a UN peacekeeping operation are socio-economic recovery and humanitarian assistance.  These are marked as supporting roles, since the Department of Peacekeeping Operations aims to support the work of other organisations and UN departments in these areas, rather than become directly involved.  For completeness, these final two activities will be examined for any value that they could bring in relation to cyber warfare.

\subsection{Socio-Economic Recovery}
Long term social and economic recovery is important for long lasting peace and ultimately for the future prosperity of a nation.  The DPKO is open about lacking the expertise and resources for this task, and peacekeeping operations are often not mandated to become involved.  Despite this, peacekeeping operations are often able to support the process where necessary through facilitating efforts and acting as a conduit between multiple organisations.  

It is proposed that there is one potential area where cyber peacekeeping could bring value.  Public trust in financial systems is an essential component of any economy~\cite{foresight2012}, and should that trust be damaged by cyber warfare, there is potential for a threat to peace and security.  In such a scenario, it is arguable that cyber peacekeepers could play a role in securing it, thereby increasing public confidence in the nation's financial system.  Despite this potential, the stated approach of this research is to respect existing peacekeeping doctrine.  It is therefore arguable that such a task would be better suited to other organisations, rather than cyber peacekeepers, especially when considering the long term nature of economic recovery and the cost of a cyber peacekeeping operation.  In line with existing practice, it is therefore concluded that the primary responsibility for socio-economic recovery remains with other organisations and departments, but that cyber peacekeeping can provide technical support where necessary.

\subsection{Humanitarian Assistance}
Humanitarian assistance is primarily a kinetic activity, aimed at providing food, shelter and other essential services to people in need.  It is difficult to propose a role that cyber peacekeepers could play in this activity, and therefore its value towards bringing peace and security is low.  It can be argued that enabling (and reducing the need for) humanitarian efforts via ensuring functioning public services e.g. power, clean water etc. is valuable, but this result is a by-product of other cyber peacekeeping tasks such as a buffer zone, malware action and restoration of state authority.  It is therefore concluded that cyber peacekeeping has no direct value towards humanitarian assistance, only an indirect one through the performance of other activities.

\section{Future Work}\label{futurework}
In this article we have explored the concept of cyber peacekeeping and examined each existing activity for value and feasibility in a post-cyber warfare context.  While we have proposed an initial short list of activities which would be both practical to perform and valuable towards maintaining peace, we have also discovered new areas of interest which are worth studying further.

\subsection{Further development of activities}
This article has briefly covered each UN peacekeeping activity and has reached an initial conclusion on which hold the potential to be valuable and feasible in recovering from cyber warfare and which would not.  This is effectively a "first pass" to shortlist those with the most potential.  In the longer term, further work is needed to go deeper into each activity and fully explore how it could be technically performed.  This work will need to draw from multiple disciplines.  For example, with regards to protecting people's right to privacy there are many research fields to examine and draw expertise from.  These include technical fields such as advances in privacy-preserving technology~\cite{Ferrag2017,Mukherjee2017,Ferrag2018} and wider social and political debates such as how human rights relate to cyber space~\cite{hick2016}.

\subsection{Compliance with core UN values}
While we have considered activities according to two criteria - value and feasibility, future work must also consider how each activity would comply with the core UN values of consent, impartiality and non-use of force.  If any activity violates these values, it would not be suitable as a UN activity.

\subsection{Challenges of Cyber Peacekeeping}
Cyber peacekeeping will face a number of challenges, some of which have been already discussed such as attribution.  But we must also consider broader challenges, both technical and political.  Firstly there is the issue of resistance at all levels - local, regional and national.  While two opposing factions or nations may agree to peace at the highest level, this may not translate into automatic compliance with cyber peacekeepers at ground level, especially at sensitive sites such as CNI.  Further exploration of the potential for resistance and how to overcome it would be valuable.

A second obstacle is the technical challenges CNI presents.  Facilities such as power plants and water facilities have properties which make observation and monitoring more challenging than standard network monitoring techniques~\cite{Nicholson2014}.  The use of proprietary protocols, air-gapping and a 24/7 availability requirement means that OMR or a cyber buffer zone on these systems will require a specialised set of skills that are in high demand globally.  

This also leads to a third challenge: securing peacekeepers in the numbers required with these highly desirable skillsets will be a challenge in itself.  UN peacekeeping operations of today are carried out by a combination of full time UN staff, and contributions of troops and police from UN member states.  Potential sources of cyber peacekeepers will therefore likely include cyber military units from UN member states (e.g. US Cyber Command).  However, it will also be important to consider other sources such as cyber security experts from private industry.  A discussion on how the right expertise can be secured in the necessary numbers, and at a price that is within an operation's budget will therefore be a necessary future research topic.

It is likely that many more obstacles exist, and we recommend that each activity concluded to be valuable and feasible should be developed in more detail to discover where additional obstacles lie and how to resolve them.

\subsection{Other Peace Activities}
While this article has focused upon UN peacekeeping, it would also be prudent to explore other peace activities such as conflict prevention and peace enforcement.

Considering conflict preventing, the UN has shown a desire to be more proactive and prevent conflict before it erupts.  Preventing conflict is cheaper than attempting to manage it once it has erupted~\cite{Annan2001,Bellamy2010} and it has been argued that the UN's primary function should be to prevent violent conflict from starting, rather than attempting to restore peace in its aftermath~\cite{Annan2001}.  There is also an argument that the international community has a moral duty to prevent harm to civilians wherever possible, and not simply in the period following warfare~\cite{Bellamy2010}.  Preventive peace deployments of the past have been generally regarded as successes~\cite{Stefanova1997}.  The concept of preventive cyber deployments is therefore something which could be valuable in the future.  It is also in line the findings of our research that cyber peacekeeping will focus upon preventing cyber attacks, rather than attempting to attribute them.

It is also reasonable to consider peace enforcement.  Could the UN authorise cyber peace enforcement in order to protect civilians?  While such an event seems valuable, the feasibility would likely be questionable.  Research into cyber warfare~\cite{Robinson2014} has shown that there is still no answer to questions such as what constitutes an armed attack or what the ethical boundaries of cyber warfare are.  With the five permanent members of the Security Council all holding a veto, unanimous consensus is difficult to achieve even in the kinetic domain.  In the cyber domain, these problems are amplified to a point where it is difficult to foresee the Security Council agreeing to enforce peace based upon a cyber conflict.  In depth study of how cyber peace enforcement could work and the value it could bring would be useful.

\subsection{Cyber Peacekeeping Reservists}\label{reservists}
One of the biggest obstacles towards the feasibility of performing valuable activities was the attribution problem.  Research is ongoing in this area, but future work in the area of cyber peacekeeping reservists based at backbone providers would be beneficial to evaluate whether it would bring a solution quicker.  The rationale behind the concept is based upon research by Wheeler and Larsen~\cite{Wheeler2003}, who state that solving the attribution problem requires prepositioned trust between backbone providers, enabling them to work together and trace attacks when required.  This was shown to be difficult, with Wheeler and Larsen~\cite{Wheeler2003} stating that ``to be effective many attribution techniques require some sort of cooperation by networks along the path from the attacker to the victim. Gaining such trust, unfortunately, can be very difficult''~\cite{Wheeler2003}.  A second identified problem is that of funding, with Wheeler and Larsen~\cite{Wheeler2003} noting that attribution technology will have a financial cost.  Without a clear business case to invest in attribution, commercial network operators find no reason to purchase it.

It was proposed that the UN is an ideal organisation from which prepositioning of trust and funding of attribution technology could take place.  The UN would approach network owners and request for one or two staff members to become cyber peacekeeping reservists.  These people would not leave the organisation, but would be able to ``activate'' when required and cooperate with other reservists at other networks to perform attribution together.  The UN would fund the required technology, since it is envisioned that this cost would be insignificant considering the value that reliable attribution of cyber attacks would bring to international peace and security.  The following research questions for future work are therefore proposed:

\begin{itemize}
	\item Would internet backbone organisations be willing to join the initiative? Potential incentives to join are compensation, free training for the reservists, and the technology being supplied for free.
	\item What would the cost of the scheme be (purchase of technology, maintenance, training etc.) and would the UN be willing to fund it?
	\item Could the scheme bring in additional funding e.g. from nations using the service to trace cyber attacks outside of a warfare context?
	\item What attribution technology is available for use in the scheme?
	\item Who should be selected as a reservist and what issues may arise? (Leaving the organisation, unavailability etc.)
	\item How do reservists come together to coordinate a trace back and what is practically required to do so?
	\item Are there any existing models which it could build upon? (e.g. Estonian cyber defence unit)
	\item Is the final result effective?  This could involve trials of the scheme.
	\item Could the scheme address more than just attribution? E.g. The ability to counter increasingly massive distributed denial of service attacks through the quick cooperation of multiple backbone organisations.
\end{itemize}

Answering these questions would provide a significant contribution towards the concept of cyber peacekeeping reservists, and would help to establish if the scheme was viable.  This would have a significant impact on global peace and security, since it would improve the potential for cyber attacks to be properly attributed and potentially provide further benefits such as an internationally coordinated response to distributed denial of service attacks.

\section{Conclusion}\label{conclusion}
Research into conducting and understanding cyber warfare is extensive and wide ranging~\cite{Robinson2014}, yet research into restoring peace after cyber warfare is sorely lacking.  In this article, we have begun to address this gap in research.  We argued that as cyber warfare becomes an increasing part of wider conflict, peacekeeping organisations such as the United Nations will find it necessary to perform cyber peacekeeping.

\begin{figure}[ht]
	\centering
	\includegraphics[width=\columnwidth]{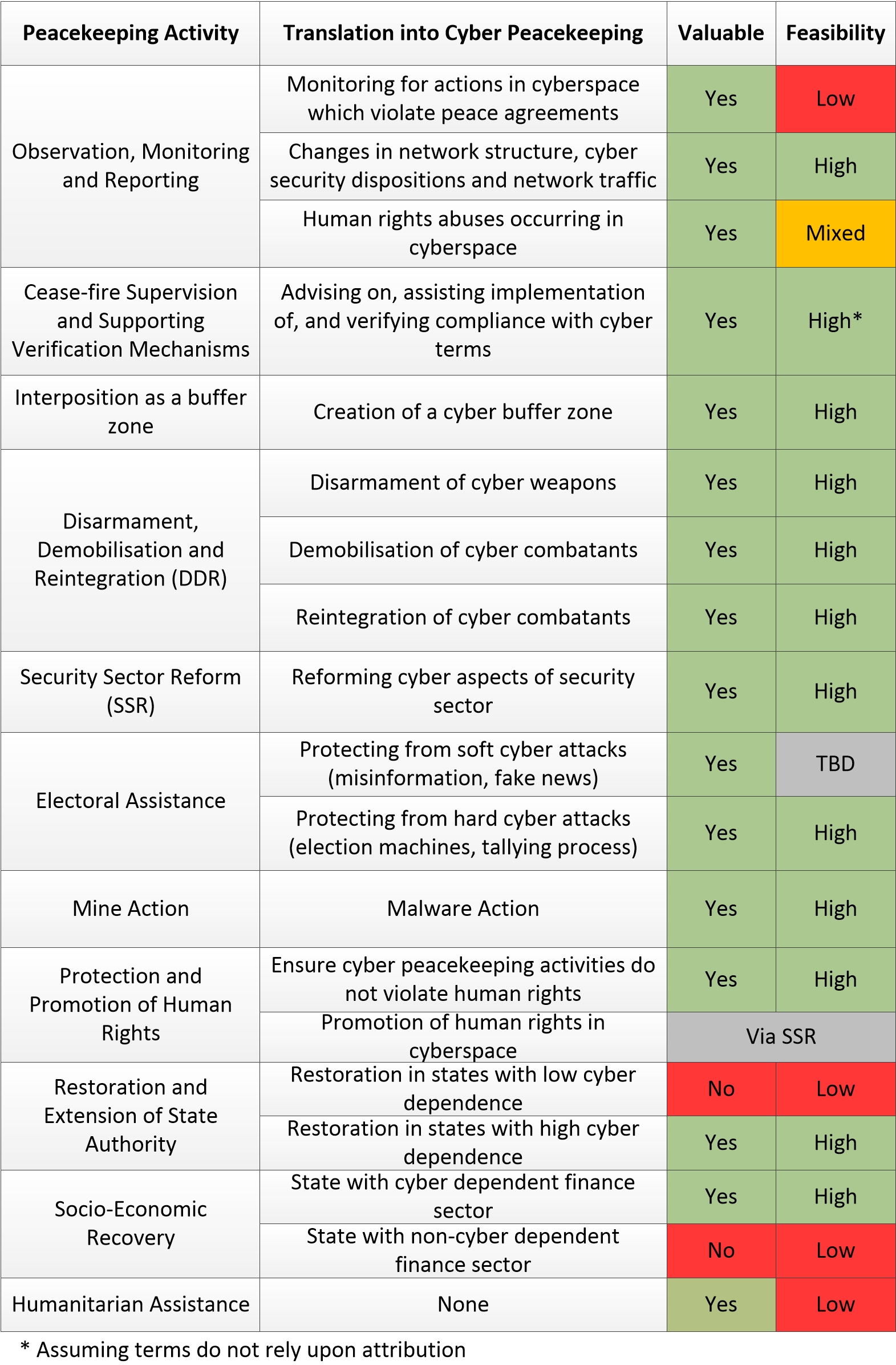}
	\caption{Summary of cyber peacekeeping activities}
	\label{fig:activitysummary}
\end{figure}

We offered a definition of cyber peacekeeping followed by an analysis of existing work on the topic.  We opted to approach the topic in a manner that respects and builds upon existing UN peacekeeping doctrine.  We justified this approach, showing that it would bring a number of benefits.  We then examined all UN peacekeeping activities and discussed the value and feasibility of performing each one in a post-cyber warfare context.  The results are summarised in figure~\ref{fig:activitysummary}, and show that there are valuable and feasible activities that cyber peacekeepers could perform.  

We conclude that while cyber peacekeeping is not necessarily needed today, it will be required in the near future as cyber warfare becomes more commonplace.  Organisations such as the UN will find it an increasing necessity to operate in cyberspace in order to maintain peace.  This article has only begun the process of defining this new role, and future work must build upon these foundations so that peacekeeping organisations can be best prepared to maintain peace in the cyber domain.

\ifCLASSOPTIONcaptionsoff
  \newpage
\fi

\bibliographystyle{ieeetr}
\balance
\bibliography{refs}

\end{document}